\documentclass[aps,prd,nofootinbib,superscriptaddress,preprintnumbers,amsmath,amssymb,latexsym,array,enumerate,letterpaper,twocolumn]{revtex4}
\usepackage{epsfig}
\usepackage[colorlinks=true,urlcolor=brown,linkcolor=blue,citecolor=magenta]{hyperref}
\usepackage{breakurl}
\usepackage{lineno}
\usepackage{bm,color,xcolor}
\usepackage{slashed} 
\usepackage{graphicx}
\usepackage{soul} 
\usepackage{multirow}
\usepackage{comment}
\usepackage{epstopdf} 
\usepackage{mathtools}
\usepackage{appendix}
\usepackage{url}
\usepackage{doi}
\usepackage[english]{babel}

\makeatletter
\def\simgt{\mathrel{\lower2.5pt\vbox{\lineskip=0pt\baselineskip=0pt
           \hbox{$>$}\hbox{$\sim$}}}}
\def\simlt{\mathrel{\lower2.5pt\vbox{\lineskip=0pt\baselineskip=0pt
           \hbox{$<$}\hbox{$\sim$}}}}
\makeatother

\newcommand{\be}{\begin{equation}}
\newcommand{\ee}{\end{equation}}
\newcommand{\bea}{\begin{eqnarray}}
\newcommand{\eea}{\end{eqnarray}}
\newcommand{\beq}{\begin{eqnarray}}
\newcommand{\eeq}{\end{eqnarray}}
\newcommand{\Fig}[1]{Fig.~\ref{#1}}
\newcommand{\Eq}[1]{Eq.~(\ref{#1})}

\newcommand{\Sec}[1]{Sec.~\ref{#1}}

\newcommand{\MeV}{\textrm{MeV}}
\newcommand{\MBH}{M_\textrm{PBH}}

\newcommand{\fPBH}{f_\textrm{PBH}}

\def\lsim{\mathrel{\rlap{\lower4pt\hbox{\hskip1pt$\sim$}}
     \raise1pt\hbox{$<$}}}         
\def\gsim{\mathrel{\rlap{\lower4pt\hbox{\hskip1pt$\sim$}}
     \raise1pt\hbox{$>$}}}         

\begin{document}

\widetext

\title{Detecting Axion-Like Particles with Primordial Black Holes}
\author{Kaustubh~Agashe}
\email{kagashe@umd.edu}
\affiliation{Maryland Center for Fundamental Physics, Department of Physics, University of Maryland, College Park, MD 20742, USA}

\author{Jae~Hyeok~Chang} 
\email{jaechang@umd.edu}
\affiliation{Maryland Center for Fundamental Physics, Department of Physics, University of Maryland, College Park, MD 20742, USA}
\affiliation{Department of Physics and Astronomy, Johns Hopkins University, Baltimore, MD 21218, USA}

\author{Steven~J.~Clark} 
\email{sclark@hood.edu}
\affiliation{Department of Physics, Brown University, Providence, RI 02912-1843, USA}
\affiliation{Brown Theoretical Physics Center, Brown University, Providence, RI 02912-1843, USA}
\affiliation{Hood College, Frederick, MD 21701, USA}

\author{Bhaskar~Dutta} 
\email{dutta@tamu.edu}
\affiliation{Mitchell Institute for Fundamental Physics and Astronomy, Department of Physics and Astronomy, Texas A\&M University, College Station, TX 77845, USA}

\author{Yuhsin~Tsai} 
\email{ytsai3@nd.edu}
\affiliation{Department of Physics, University of Notre Dame, IN 46556, USA}

\author{Tao~Xu} 
\email{tao.xu@ou.edu}
\affiliation{Department of Physics and Astronomy, University of Oklahoma, Norman, OK 73019, USA}
\affiliation{Racah Institute of Physics, Hebrew University of Jerusalem, Jerusalem 91904, Israel}

\preprint{
\begin{minipage}{5cm}
\begin{flushright}
UMD-PP-022-12 \\
MI-HET-792
 \end{flushright}
\end{minipage}
}

\begin{abstract}
Future gamma-ray experiments, such as the e-ASTROGAM and AMEGO telescopes, can detect the Hawking radiation of photons from primordial black holes (PBHs) if they make up a fraction or all of dark matter. PBHs can analogously also Hawking radiate new particles, which is especially interesting if these particles are mostly secluded from the Standard Model (SM) sector, since they might therefore be less accessible otherwise. A well-motivated example of this type is axion-like particles (ALPs) with a tiny coupling to photons. We assume that the ALPs produced by PBHs decay into photons well before reaching the earth, so these will augment the photons directly radiated by the PBHs. Remarkably, we find that the peaks in the energy  distributions of ALPs produced from PBHs are different than the corresponding ones for Hawking radiated photons due to the spin-dependent greybody factor. Therefore, we demonstrate that this process will in fact distinctively modify the PBHs' gamma-ray spectrum relative to the SM prediction. We use monochromatic asteroid-mass PBHs as an example to show that e-ASTROGAM can observe the PBH-produced ALP gamma-ray signal (for masses up to $\sim 60$~MeV) and further distinguish it from Hawking radiation without ALPs. By measuring the gamma-ray signals, e-ASTROGAM can thereby probe yet unexplored parameters in the ALP mass and photon coupling.

\end{abstract}

\maketitle

\section{Introduction}

Future satellite telescopes, like the proposed e-ASTROGAM~\cite{e-ASTROGAM:2016bph} and AMEGO~\cite{AMEGO:2019gny} experiments, will play a vital role in multimessenger astrophysics and cover the energy gap in the current gamma-ray observations between order $0.1$ to $10$~MeV scales. This energy window includes motivated target signals from beyond the Standard Model (BSM) physics, such as the gamma-rays produced from dark matter (DM) annihilation~\cite{Boehm:2002yz, Beacom:2004pe, Finkbeiner:2007kk, Essig:2009jx, Essig:2013goa, Boddy:2015efa, Laha:2020ivk}, from the Hawking radiation of primordial black holes (PBHs)~\cite{Carr:2009jm, Laha:2019ssq, Carr:2020gox, Coogan:2020tuf, Laha:2020ivk, Ray:2021mxu,Keith:2022sow, Tan:2022lbm, Caputo:2022dkz}, or from the decay of axion-like particles (ALPs) produced in the early universe~\cite{Caputo:2022dkz}. 
If more than one type of BSM physics exists in nature, the new physics objects can also couple to each other through SM or BSM interactions, or at a minimum via gravity. A careful measurement of the gamma-ray spectrum in these experiments may simultaneously identify signals with more than one BSM origin.

In this work, we investigate the possibility of using the e-ASTROGAM experiment to identify the Galactic Center gamma-ray signal from PBHs. The signal is composed of both direct PBH production of photons and indirect production from ALPs Hawking radiated by the PBHs. These ALPs subsequently decay into photons well before reaching the earth, producing the indirect secondary signal. In particular, we study the Hawking radiation of PBHs with asteroid-scale mass $M_{\rm PBH}\sim 10^{15-17}$~g that emit with Hawking temperatures $T_H\approx(10^{16}{\rm g}/M_{\rm PBH})$~MeV $\sim \mathcal{O}(0.1-10)$ MeV and have a lifetime $\tau\approx 10^{5}(M_{\rm PBH}/10^{16}{\rm g})^3$~Gyrs comparable to the age of the universe if $M_{\rm PBH}\lsim 10^{15}$~g. We map out the ALP and PBH parameters for observing the galactic gamma-ray signal at e-ASTROGAM, and further discuss the possibility of distinguishing signals from PBH with and without ALP production.
 
Before discussing our work, we provide a quick review of the components in our analysis, starting with PBHs.
The PBH has long been considered a plausible candidate for all or a fraction of cold dark matter (CDM)~\cite{Carr:2016drx}. There are many proposals for producing PBHs in the early universe, such as a cosmological scenario that produces an order one density contrast in the early universe ~\cite{Carr:1975qj,Ivanov:1994pa,Garcia-Bellido:1996mdl,Silk:1986vc,Kawasaki:1997ju,Yokoyama:1995ex,Pi:2017gih,Hertzberg:2017dkh, Ozsoy:2018flq, Cicoli:2018asa,Ashoorioon:2019xqc}, first order phase transitions~\cite{Hawking:1982ga,Crawford:1982yz,Kodama:1982sf,Moss:1994pi,Freivogel:2007fx,Johnson:2011wt,Kusenko:2020pcg,Ashoorioon:2020hln,Baker:2021nyl, Baker:2021sno, Kawana:2021tde,Huang:2022him, Lu:2022paj,Ashoorioon:2022raz}, the dynamics of scalar field fragmentations~\cite{Cotner:2016cvr,Cotner:2017tir,Cotner:2018vug,Cotner:2019ykd}, collapse of cosmic strings~\cite{Hawking:1987bn, Polnarev:1988dh,MacGibbon:1997pu, Brandenberger:2021zvn} or domain walls~\cite{Rubin:2000dq,Rubin:2001yw}, and holographic cosmology~\cite{Banks:2018ypk, Banks:2020dgx, Banks:2021lai}. Hawking radiation produces BSM particles with rates that only depend on particles' masses and spins regardless of their coupling to the SM sector. Therefore, PBHs are a source for producing BSM particles even if their relic abundance is negligible. Many studies discuss the PBH production of DM or dark radiation~\cite{Bell:1998jk, Allahverdi:2017sks, Lennon:2017tqq, Hooper:2019gtx, Gondolo:2020uqv, Cheek:2021odj, Cheek:2021cfe, Hooper:2020evu, Arbey:2021ysg, Sandick:2021gew, Masina:2021zpu,Cheek:2022dbx}, ALP~\cite{Schiavone:2021imu,Bernal:2021yyb, Mazde:2022sdx,Li:2022mcf}, or the baryon asymmetry~\cite{Zeldovich:1976vw, Carr:1976zz, Toussaint:1978br, Turner:1979bt, Grillo:1980rt, Baumann:2007yr, Fujita:2014hha, Hook:2014mla,Hamada:2016jnq, Morrison:2018xla, Hooper:2020otu, Bernal:2022pue, Gehrman:2022imk} before Big-bang Nucleosynthesis (BBN), i.e., in the relatively early universe. Our focus instead will be on the PBH production of new particles at a low redshift~\cite{Baker:2021btk,Calabrese:2021src,Calza:2021czr,Li:2022jxo,Calabrese:2022rfa,Li:2022xqh,Baker:2022rkn}, so that the relevant experiments can detect PBH signals both from the SM and the BSM particles that they produce {\em currently}.

Previous analyses from the COMPTEL~\cite{Kappadath:1998PhDT} and Fermi-LAT~\cite{Fermi-LAT:2009ihh} experiments set upper bounds on the 
direct
gamma-ray flux 
from PBHs. The better sensitivity of the e-ASTROGAM experiment opens up a new window for PBH observation between $E_\gamma\approx 0.1-10$~MeV and flux $E^2_\gamma d\Phi/dE_\gamma\approx 10^{-6}-10^{-5}$~MeVcm$^{-2}$s$^{-1}$, which corresponds to signals from PBHs with a monochromatic asteroid-scale mass and energy density fraction $f_{\rm BH}$ (of the Milky Way DM density) that satisfies $10^{-6}\lsim f_{\rm PBH}(10^{16}\,{\rm g}/M_{\rm PBH})^3\lsim 10^{-4}$~\cite{Coogan:2020tuf}.

\begin{figure}[t!]
\centering
\includegraphics[scale=0.4]{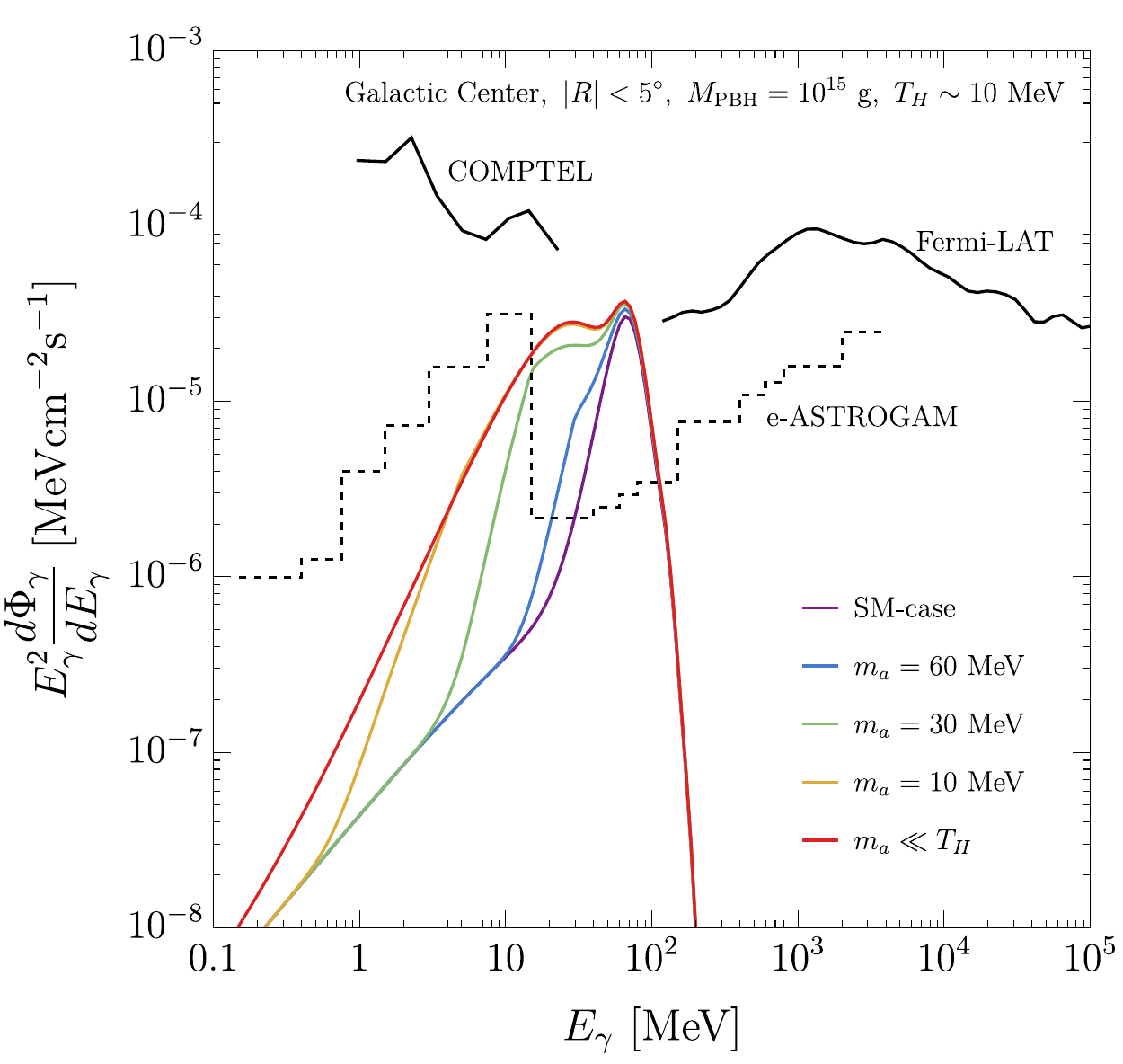}
\caption{An example of the gamma-ray spectrum from Hawking radiation with the SM-case~(Purple) and the ALP-case with $m_a=60~{\rm MeV}$~(Blue), $m_a=30~{\rm MeV}$~(Green), $m_a=10~{\rm MeV}$~(Yellow), $m_a\ll T_H$~(Red). The PBH mass is $M_{\rm PBH}=10^{15} {\rm g}$, corresponding to $T_H\approx10$~MeV, and the energy density fraction of DM is $f_{\rm PBH}=10^{-8}$. In this figure, we assume the ALP decay length is much shorter than the distance to the sources (PBHs) and consider its decay to be prompt (instantaneous after creation) when calculating the ALP-case flux. The ROI is chosen to be the Galactic Center with $|R|\leq 5^{\circ}$. Experimental constraints and future sensitivity are shown in black for COMPTEL~(Solid), Fermi-LAT~(Solid), and e-ASTROGAM~(Dashed).} 
\label{fig:ALPspectrum}
\end{figure}

In this work, we extend the target for the gamma-ray search of BSM physics from PBHs alone to new particles produced by the PBHs, 
focusing on ALPs as an illustration. 
Namely, we study the signal in the ``ALP-case", which refers to the PBH scenario that produces ALPs from Hawking radiation, 
followed by their prompt decay into photons,
thus modulating the direct photon production by PBHs.
Then, we compare the signal to the ``SM-case", which refers to the PBH scenarios with only SM particles. Because the observation of the galactic gamma-ray signal sets a stronger bound on the PBH abundance than the search of dwarf spheroidal galaxies~\cite{Coogan:2020tuf}, we focus on the Milky Way galactic signal for the ALP-case. 

ALPs are a widely studied subject in particle physics and cosmology, where the axion field was originally proposed to solve the strong CP problem in QCD~\cite{PhysRevLett.38.1440, PhysRevD.16.1791} and later realized to be a viable DM candidate~\cite{ABBOTT1983133,PRESKILL1983127,DINE1983137, Co:2017mop}. The ALP generalizes the phenomenology of the QCD axion without a necessary connection to the strong CP problem, but it can help to address other physics questions such as cosmic inflation~\cite{Freese:1990rb, Kim:2004rp, Pajer:2013fsa}, hierarchy problems~\cite{Graham:2015cka, Gupta:2015uea, Choi:2015fiu}, or serve as a benchmark target for DM searches~\cite{Sikivie:2020zpn,Bauer:2017ris, Bauer:2018uxu, Adams:2022pbo}. Some earlier experiments and astrophysical and cosmological studies have also excluded part of the parameter space for the axion mass and coupling, see~\cite{Irastorza:2018dyq, DiLuzio:2020wdo, Choi:2020rgn} for recent reviews.
\begin{figure}[t!]
\centering
\includegraphics[scale=0.4]{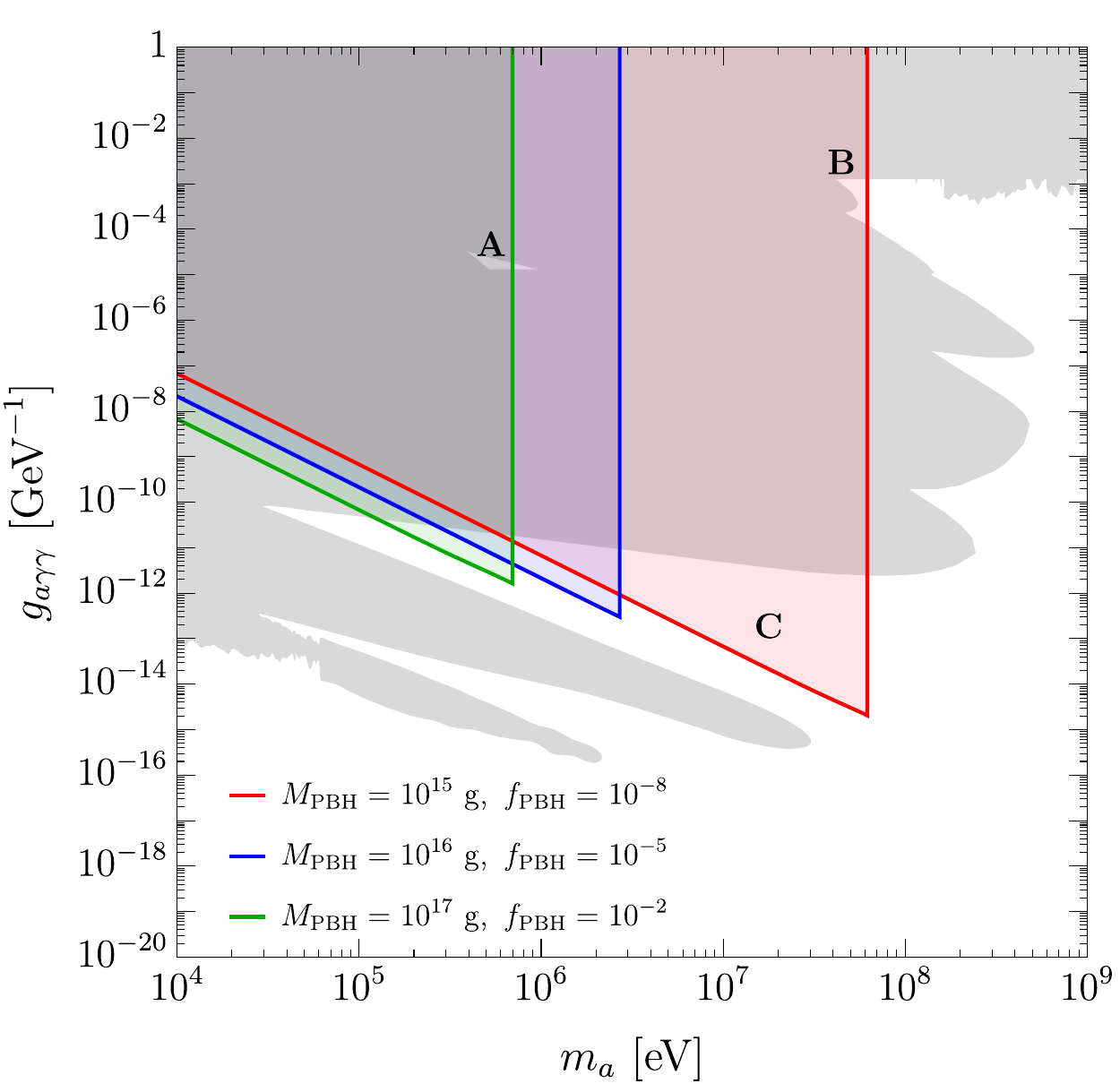}
\caption{The ALP parameter space that can be probed by PBHs. The PBH parameters are chosen as $M_{\rm PBH}=10^{15}~{\rm g}$, $f_{\rm PBH}=10^{-8}$ (Red); $M_{\rm PBH}=10^{16}~{\rm g}$, $f_{\rm PBH}=10^{-5}$ (Blue); $M_{\rm PBH}=10^{17}~{\rm g}$, $f_{\rm PBH}=10^{-2}$ (Green). Existing bounds (Grey) are taken from~\cite{AxionLimits,Ayala:2014pea,2015JCAP...10..015V, Knapen:2016moh, Belle-II:2020jti, Lucente:2020whw, Dolan:2021rya, Dessert:2021bkv, Caputo:2022mah, Dessert:2022yqq, Kling:2022ehv, DeRocco:2022jyq, Balazs:2022tjl, Dolan:2022kul, Langhoff:2022bij, BESIII:2022rzz}.}
\label{fig:ALPparam}
\end{figure}
If the ALP is indeed responsible for one of the physics puzzles described above, the existence of PBHs guarantees ALP production when the process is kinetically allowed. We consider the ALP scenario with the Lagrangian
\bea
\mathcal{L}_{a\gamma\gamma} \supset	\frac{1}{2}\partial_\mu a \, \partial^\mu a -\frac{1}{2}m^{2}_{a} a^{2}_{a} + \frac{g_{a\gamma\gamma}}{4} a F_{\mu\nu}\tilde{F}^{\mu\nu}\,.
\eea
The ALP can decay into two photons with the $a F_{\mu\nu}\tilde{F}^{\mu\nu}$ coupling. Here we assume the di-photon channel dominates the ALP decay for the mass $m_a < 100$~MeV \footnote{
For the PBH masses we consider, ALPs with $m_a>100$~MeV have negligible production rates. For $m_a<100~{\rm MeV}<2m_{\pi}$, decay into SM hadrons is kinematically forbidden.} considered in this work. This assumption is realized in axion models where the SM leptons do not carry the Peccei-Quinn (PQ) charge, such as the well-studied KSVZ model where axion couples to SM through heavy vector-like quarks~\cite{SREDNICKI1985689, Kim:2008hd}. In this case, the ALP decay width in the rest frame is
\begin{equation}
\Gamma_a=\frac{g^2_{a\gamma\gamma}~m^3_a}{64\pi}.
\end{equation}

We can then 
summarize our results of working out the ``interaction'' between PBHs and ALPs as follows.
Since PBHs produce ALPs with a nearly thermal energy distribution $E_a$, after traveling distance $D$ from the PBH, the produced ALPs decay into photons with a probability
\begin{equation}\label{eq:Padecay}
P_{a, \textrm{decay}}(E_a, D)=1-\exp\left(-D~\Gamma_a \frac{m_a}{\sqrt{E^2_a-m^2_a}}\right).
\end{equation}
The ALP-case, therefore, has a gamma-ray spectrum both from the direct PBH production and a secondary emission from the ALP decay. In Fig.~\ref{fig:ALPspectrum}, we show the photon spectrum in the ALP-case for $M_{\rm PBH}=10^{15}$~g, $f_{\rm PBH}=10^{-8}$, and different ALP masses. In the example, the Hawking temperature $T_H\approx 10$~MeV generates a primary gamma-ray spectrum peaked around  $10 \; T_H$~\cite{Agashe:2022jgk}. In general, due to the spin-0 greybody factor, $E_\gamma<E_a$ from the ALP decay, and the suppressed production of ALPs much heavier than the primary photon peak, the visible ALP peak is always on the left side of the primary peak. The ALP spectrum has a different peak location and spectral shape than the photon spectrum in the SM-case, and the double-peak feature of the spectral shape makes it possible to distinguish between the signals for the ALP- and SM-cases.  
We leave the details of the spectrum calculation to Sec.~\ref{sec:HawkingRadiation}. 

For claiming the observation of the ALP signal, since we do not know {\it a priori} the PBH mass and abundance, we need to be able to distinguish the photon spectrum in Fig.~\ref{fig:ALPspectrum} from the SM-case with arbitrary PBH parameters. We conduct a less ambitious study in this work by assuming non-rotating PBH with a monochromatic mass spectrum. With the expected e-ASTROGAM sensitivity given in~\cite{AMEGO:2019gny}, we calculate the region of $m_a$ and $f_{\rm PBH}$ for a given $M_{\rm PBH}$ that allows the e-ASTROGAM to differentiate signals in the ALP-case from the SM-case. 

Fig.~\ref{fig:ALPparam} summarizes regions of the ALP parameters we identify that allow  $3\sigma$ e-ASTROGAM differentiation between the ALP- and SM-case signals. The lower boundaries of each colored region give the minimum $g_{a\gamma\gamma}$ for ALPs produced at Galactic Center from PBH to have $P_{a, \rm decay}=0.99$ decay before reaching the earth.\footnote{$P_{a, \rm decay}=0.99$ was taken as an estimate for the maximum sensitivity as it approximates a near complete ALP decay. This corresponds to where the later used gamma-ray analysis can be conducted irrespective of $g_{a\gamma\gamma}$ and should be viewed as a conservative estimate for the region. A more careful analysis would be able to slightly extend the region but is beyond the scope of this work.} We consider the $E_a$ distribution from the Hawking radiation in the probability calculation described in Sec.~\ref{sec:HawkingRadiation}. The right boundary of each colored region indicates the maximum ALP mass that can be produced and generate distinct enough signals between the two cases. We explain the details of the calculation in Sec.~\ref{sec.liekli}.

Fig.~\ref{fig:ALPparam} also shows the existing exclusion bounds (light grey) from various collider, astrophysical, and cosmological analyses. To enlarge the allowed ALP parameter space for the indirect detection signal, we already assume a low reheating temperature $T_{\rm reh}=5$~MeV that weakens the BBN+$\Delta N_{\rm eff}$ bound \cite{Balazs:2022tjl,Langhoff:2022bij}. Our result shows that the existing bounds have tightly constrained
the ALP parameter space for producing a distinguishable galactic center gamma-ray signal from the PBH emission, besides three regions labeled in the plot. Parameters around the so-called cosmological triangle region at point {\bf A} can exist either with the presence of $\Delta N_{\rm eff}$, a non-vanishing neutrino chemical potential, or a lower reheating temperature~\cite{Depta:2020wmr, Depta:2020zbh}\footnote{Ref.~\cite{Caputo:2021rux} shows that the measurement of the explosion energy of SN1987A is in severe tension to ALP around the cosmological triangle unless the star cooling process is significantly different from the standard picture.}. For PBH mass around $10^{15}$~g, the corresponding Hawking temperature is high enough to produce ALPs with $m_a\approx 40-60$~MeV. The ALPs in the region around {\bf B} permit distinctive enough gamma-ray signals compared to the SM-case. However, the ALPs need to have dominant decay into photons even if the $a\to e^+e^-$ decay is kinematically allowed. As mentioned above, the dominant photon decay can be achieved in models like the KSVZ scenario where the SM leptons do not carry the PQ charge. The allowed parameter space in the {\bf C} region remains open for the ALP-case signal if the SM particles get populated only starting from a low temperature $\approx5$~MeV. A higher reheating temperature of the SM sector can thermally produce relativistic ALPs and therefore generate $\Delta N_{\rm eff}$ that violates the BBN constraint\footnote{Even with a late reheating into the SM particles, PBH production can still happen from, e.g., the gravitational collapse of a large primordial curvature perturbation.  Before the reheating process, the SM-neutral reheaton that later decays into SM particles can inherit the curvature perturbation, making some Hubble patches collapse into black holes right after the horizon re-entry.}. For the following study, we assume the ALP with a given $m_a$ has a value of $g_{a\gamma\gamma}$ that is outside of the excluded (grey) region but inside the colored (green, blue, red) region (making the signal observable) and will not specify the coupling again.

The format of the paper is as follows. In Sec.~\ref{sec:HawkingRadiation}, we review the calculation of gamma-ray and ALP production from Hawking radiation. We calculate the galactic gamma-ray signal from the ALP-case by including the proper $J$-factor integral. In Sec.~\ref{sec.liekli}, we conduct a likelihood  analysis for distinguishing the ALP-case signal from the SM-case signal. We conclude in Sec.~\ref{sec.conclusions}.

\section{Hawking Radiation from Primordial Black Holes}\label{sec:HawkingRadiation}

A black hole is expected to emit particles constantly near its event horizon, and this phenomenon is called Hawking radiation~\cite{Hawking:1974rv}. In this section, we review particle spectra of Hawking radiation, mostly following \cite{Coogan:2020tuf}. Particles directly produced from black holes are called primary particles, while particles from the result of interactions of primary particles are called secondary particles. The number of produced primary particle $i$ per unit time per unit energy from a black hole with mass $M$ is given by~\cite{Hawking:1974rv, Page:1976df, MacGibbon:1990zk}
\begin{equation}
    \frac{\partial N_{i,\textrm{primary}}}{\partial E_i \partial t} = \frac{g_i}{2 \pi}\frac{\Gamma_i(E_i,M,\mu_i)}{e^{E_i/T_H}\pm 1},
\end{equation}   
where $\mu_i$ and $g_i$ are the mass and the degree of freedom of particle $i$, $\Gamma_i$ is the greybody factor, $T_H = 1/(8 \pi G M)$ is the Hawking temperature, and the plus and minus signs correspond to whether the produced particle is a fermion or a boson, respectively. The spin-dependant greybody factor approaches the geometrical optics limit $\Gamma_i = 27 G^2 M^2 E^2_i$ for high energies. We use \textbf{BlackHawk} package~\cite{Arbey:2019mbc, Arbey:2021mbl} to get the greybody factor $\Gamma_i$ of non-rotating PBHs. The particle rest mass cuts the evaporation spectrum to $E_i \geq \mu_i$. Note that the greybody factor used in \textbf{BlackHawk} assumes massless particle production, but this should have a minimal effect on our result except for $m_a \gg T_H$.

We are interested in the photon spectrum of Hawking radiation, which includes the primary spectrum and the secondary spectrum from decay and final state radiation (FSR) of primary particles.
\begin{eqnarray}
    \frac{\partial N_{\gamma,\textrm{tot}}}{\partial E_\gamma \partial t} &=& \frac{\partial N_{\gamma,\textrm{primary}}}{\partial E_\gamma \partial t} \nonumber\\
    &&+ \sum_{i=\pi^0,a} \int d E_i 2\frac{\partial N_{i,\textrm{primary}}}{\partial E_i \partial t} \frac{d N_{i,\textrm{decay}}}{dE_\gamma}\nonumber\\    
    &&+ \sum_{i=e^\pm,\mu^\pm,\pi^\pm} \int d E_i \frac{\partial N_{i,\textrm{primary}}}{\partial E_i \partial t} \frac{d N_{i,\textrm{FSR}}}{dE_\gamma}\nonumber\\
\end{eqnarray}
with
\begin{eqnarray}
    \frac{d N_{i,\textrm{decay}}}{dE_\gamma} &=& \frac{\Theta(E_\gamma-E_i^-) \Theta(E_i^+-E_\gamma)}{E_i^+-E_i^-}\\
    E_i^\pm &=& \frac{1}{2} \left ( E_i \pm \sqrt{E_i^2 - m_i^2} \right )
\end{eqnarray}
\begin{eqnarray}
    \frac{d N_{i,\textrm{FSR}}}{dE_\gamma} &=& \frac{\alpha}{\pi Q_i}P_{i\rightarrow i\gamma}(x) \left [\log \left (\frac{1-x}{\mu_i^2} \right ) -1 \right ]\\
    P_{i\rightarrow i\gamma}(x) &=& \begin{dcases} \frac{2(1-x)}{x}, & i=\pi^\pm \\ \frac{1+(1-x)^2}{x}, & i=\mu^\pm, e^\pm \end{dcases}
\end{eqnarray}
Note we ignore the three-body decay from $\mu^{\pm}$ and $\pi^{\pm}$. These processes are safe to be ignored because they are much heavier than the energy range we are interested.

The greybody factor for scalar particles has a peak at a smaller energy compared to vectors; the ALP-case spectrum has a ``double peak'' feature as shown in \Fig{fig:ALPspectrum}, and this is distinguishable from the SM-case spectrum.

The photon flux near the Earth is
\begin{equation}
    \frac{d \Phi_\gamma}{d E_\gamma} = \bar{J}_D \frac{\Delta \Omega}{4 \pi} \int dM \frac{f_\textrm{PBH}(M)}{M} \frac{\partial N_{\gamma,tot}}{\partial E_\gamma \partial t}.
    \label{eq:photonflux}
\end{equation}
$J_D$ is the so-called J-factor for decay, which is given by
\begin{equation}
    \bar{J}_D = \frac{1}{\Delta \Omega}\int_{\Delta \Omega} d\Omega \int_\textrm{LOS} dl \rho_\textrm{DM}.
\end{equation}
To get the J-factor, we assume the DM distribution in the Milky-Way halo follows a Navarro–Frenk–White~(NFW) profile~\cite{Navarro:1996gj} 
\bea
\rho_{\rm DM}(r)=\frac{\rho_s}{\frac{r}{r_s} \, (1+\frac{r}{r_s})^2} \Theta(r_{200}-r).
\eea
We use $r_s=11~{\rm kpc}$, $\rho_s=0.838~{\rm GeV}/{\rm cm}^3$, $r_{200}=193~{\rm kpc}$ and $r_\odot=8.122~{\rm kpc}$~\cite{2019JCAP...10..037D}. For our ROI of $|R| < 5^{\circ}$ from the Galactic Center, the J-factor is $\bar{J}_D=1.597 \times 10^{26}~{\rm MeV} {\rm cm}^{-2} {\rm sr}^{-1}$ and the angular size is $\Delta \Omega=2.39 \times 10^{-2} {\rm sr}$. 

In this study, we assume the PBH mass distribution is monochromatic, which can be produced from, for example, the collapse of Q-balls~\cite{Flores:2021jas} or first-order phase transition~\cite{Jung:2021mku}. Taking $f_{\rm PBH}(M)=f_{\rm PBH}\delta(M-M_{\rm PBH})$, \Eq{eq:photonflux} simplifies to 
\beq
    \frac{d \Phi_\gamma}{d E_\gamma} = \bar{J}_D \frac{\Delta \Omega}{4 \pi}  \frac{f_\textrm{PBH}}{M_{\rm PBH}} \frac{\partial N_{\gamma,tot}}{\partial E_\gamma \partial t}
\eeq

In order for ALPs to change the photon spectrum near the earth, they must decay before reaching the earth. The decay probability of ALPs while propagating to the earth from Galactic Center with a monochromatic PBH mass is given by
\beq
\langle P_{a, \textrm{decay}} \rangle\equiv \frac{\Phi_{a,\textrm{dec}}}{\Phi_{a,\textrm{tot}}},
\eeq
where
\begin{widetext}
\begin{eqnarray}
\Phi_{a,\textrm{tot}}&=& \int_{\Delta\Omega}\frac{d\Omega}{4\pi}\int_{\rm LOS}d\ell \int dE_{a} \frac{\fPBH \rho_{\rm DM}}{\MBH} \, \frac{\partial N_{a,\textrm{primary}}}{\partial E_{a} \partial t} ~=~ \bar{J}_D \frac{\Delta \Omega}{4 \pi}  \frac{f_\textrm{PBH}}{M_{\rm PBH}} \int dE_a \frac{\partial N_{a, \textrm{primary}}}{\partial E_a \partial t}, \,\\
\Phi_{a,\textrm{dec}}&=& \int_{\Delta\Omega}\frac{d\Omega}{4\pi}\int_{\rm LOS}d\ell \int dE_{a} \frac{\fPBH \rho_{\rm DM}}{\MBH} \, \frac{\partial N_{a,\textrm{primary}}}{\partial E_{a} \partial t} \, P_{a, \textrm{decay}}(E_a,\ell), \nonumber\\
\label{eq:probavg}
\end{eqnarray}
\end{widetext}
and $P_{a, \textrm{decay}}$ is given by \Eq{eq:Padecay}.
We require $\langle P_{a, \textrm{decay}} \rangle$ larger than 99\% to get the lower boundaries of the color curves in \Fig{fig:ALPparam}. For most of the parameter regions shown in Fig.~\ref{fig:ALPparam}, we can assume the ALP decay is prompt as $g_{a\gamma\gamma}$ is at least one order of magnitude larger than the lower boundaries. Since the ALP decay width is proportional to the square of the coupling $\Gamma_a \propto g_{a\gamma\gamma}^2$, even the decay length of ALPs in region C is much smaller than the distance between the earth and Galactic Center.

\section{Experiments and Likelihood Analysis}\label{sec.liekli}

\begin{figure*}[th]
\centering
\includegraphics[width=0.98\columnwidth]{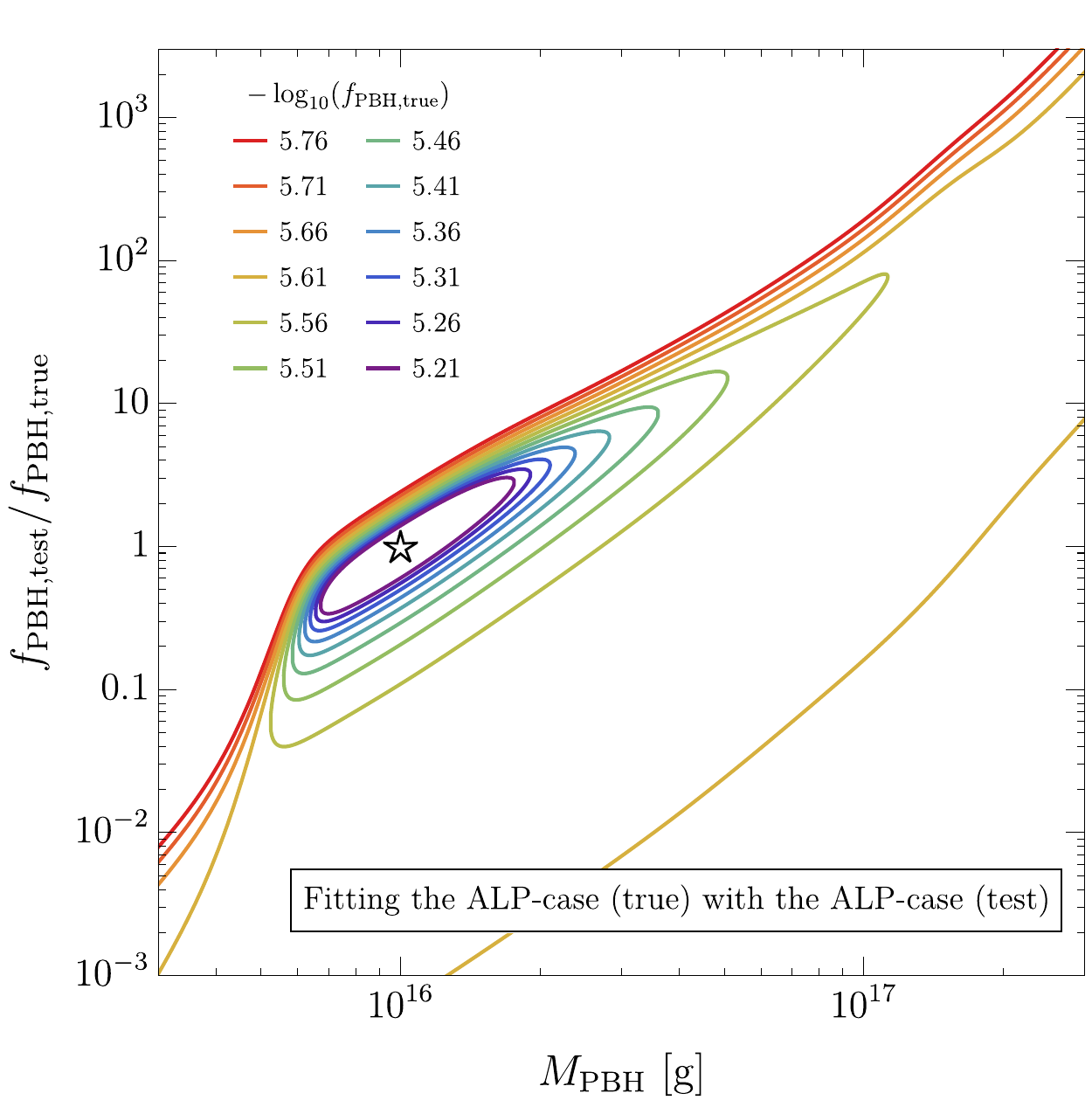}
$\quad$
\includegraphics[width=0.98\columnwidth]{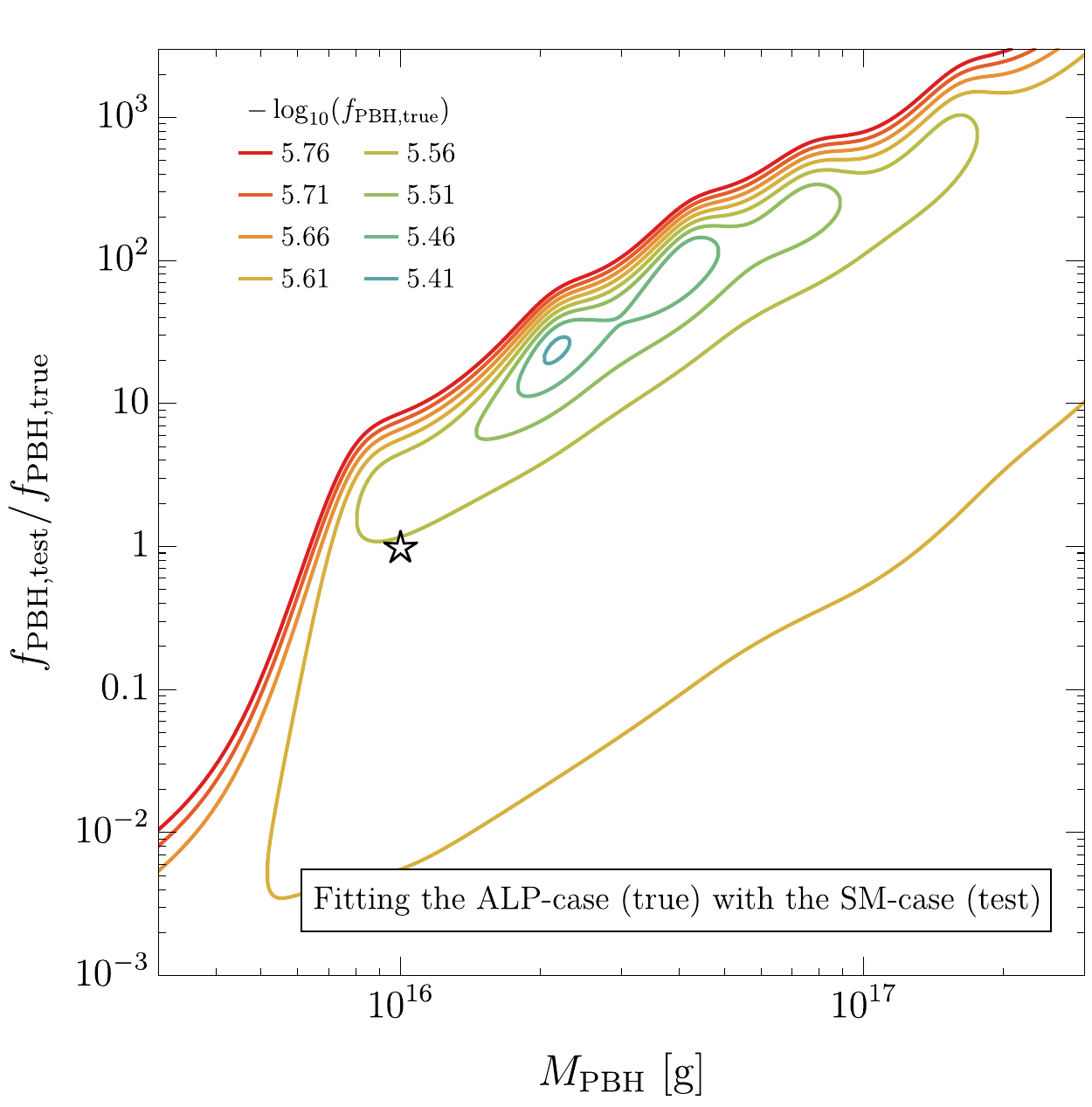}
\caption{The 3$\sigma$ contours from the Galactic Center constraining the PBH parameter space, assuming that the underlying ``true" model is the ALP-case with $m_a=10^{-1}\;\mathrm{MeV}$, $M_\mathrm{PBH}=10^{16}\;\mathrm{g}$, and various $f_\mathrm{PBH,true}$. The ``test" models are ({\it left}) the ALP-case and ({\it right}) the SM-case. The ``$\star$" indicates the location of the true model. With increasing $f_\mathrm{PBH,true}$, the contours shrink. When the test model matches the true model (as is the case in the left panel), the contours become infinitesimally small, while for a mismatched model ({\it right}), they eventually disappear with sufficient data such that no contour may be drawn. We define the values of $f_\mathrm{PBH,true}$ where the contour disappears in the right panel as ``identification of ALP'' line in \Fig{fig:PBHparamBounds}. The values of $f_\mathrm{PBH,true}$ for contours are shown in each plot.} 
\label{fig:PBHparamContour}
\end{figure*}

In order to place constraints on individual models, we use the likelihood analysis outlined in \cite{Agashe:2022jgk}. In this analysis, an assumption is made about a ``true" model that produces an observational gamma-ray signal. A test model is also chosen for comparison. The likelihood that the test model will replicate the gamma-ray signal produced by the true model follows Poisson statistics and is expressed as
\beq
\mathcal{L}= \exp{\left(\sum_i n_i \ln{\sigma_i}-\sigma_i-\ln{n_i!}\right)}\label{eq.likeli}
\eeq
where $n_i$ is the photon count of the true model (including any additional background signals) and $\sigma_i$ is the expected photon count from the test model (including background) in the $i$-th energy bin. In this work, all results assume an observation time of $10^8~\mathrm{s}$. For comparison of different models, we utilize the test statistic (TS)
\begin{equation}
    \mathrm{TS} = - 2 \ln{\left(\frac{\mathcal{L}}{\mathcal{L}_\mathrm{true}}\right)} = \Sigma^2,
    \label{eq:TS}
\end{equation}
where $\Sigma$ is the observational significance~\cite{Cowan:2010js, Rolke:2004mj, Bringmann:2012vr, Fermi-LAT:2015kyq} and $\mathcal{L}_\mathrm{true}$ is the likelihood of the true model. Here, we are assuming that the joint analysis of the gamma-ray energy bins follows a $\chi^2$ distribution, and unless otherwise stated, we take $\Sigma=3$, which corresponds to a $99\%$ discovery significance. In order to reduce complexity and statistical fluctuations in the likelihood determination, we also take the true signal as its statistical mean.

In creating the true and test signals as well as the expected background, we use the $5^\circ$ NFW Galactic Center as the source for the observational signal (see \Sec{sec:HawkingRadiation}) and use estimations of the e-ASTROGAM detector sensitivities and specifications as well as the forecasted astrophysical background \cite{e-ASTROGAM:2016bph, Agashe:2022jgk}. Please refer to Ref.~\cite{Agashe:2022jgk} for more details about the detector sensitivity and foreground used.

\subsection{Signal Detection and Distinguishability between SM and ALP}
\begin{figure*}[th]
\includegraphics[width=0.98\columnwidth]{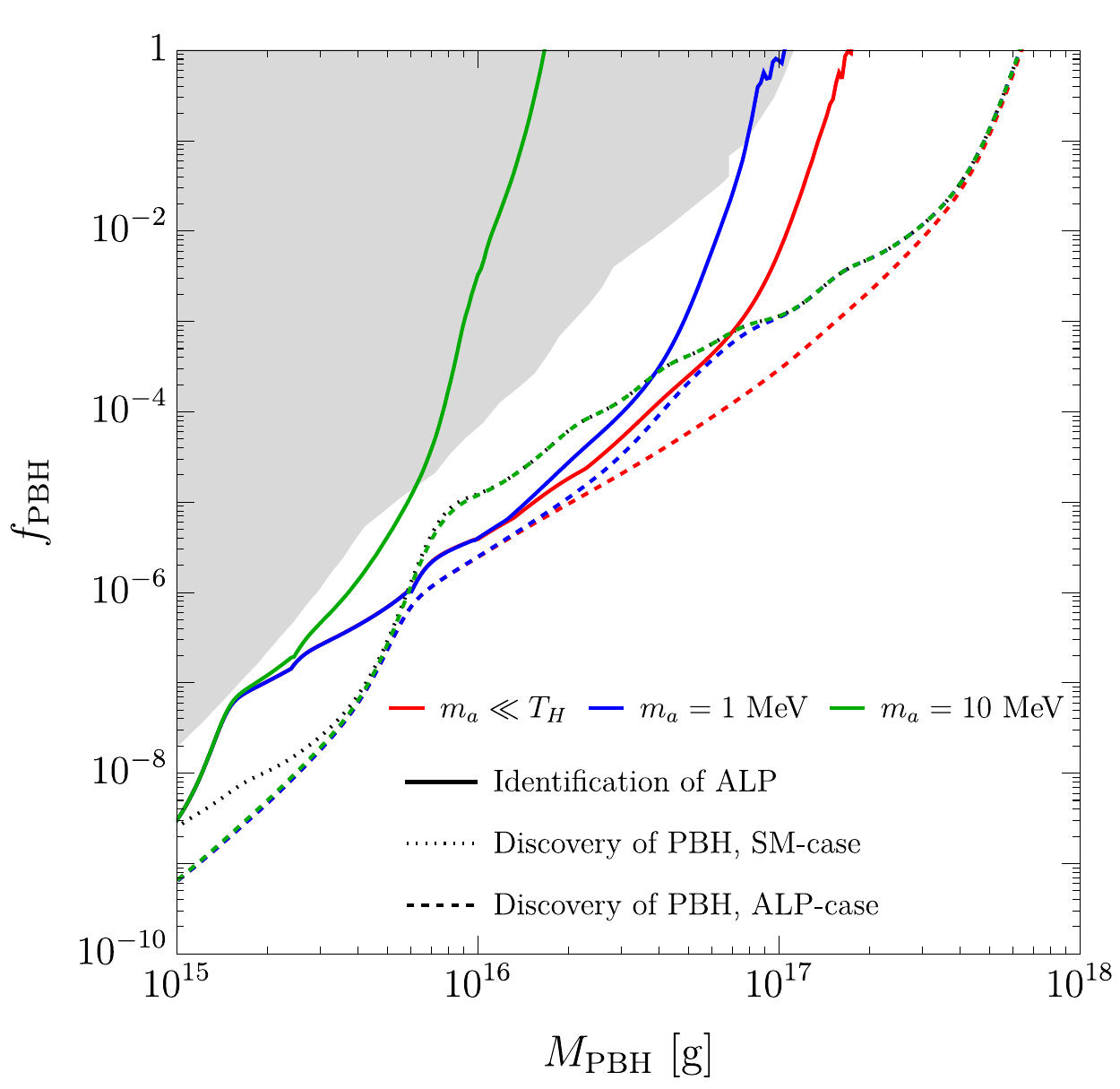}
$\quad$
\includegraphics[width=0.98\columnwidth]{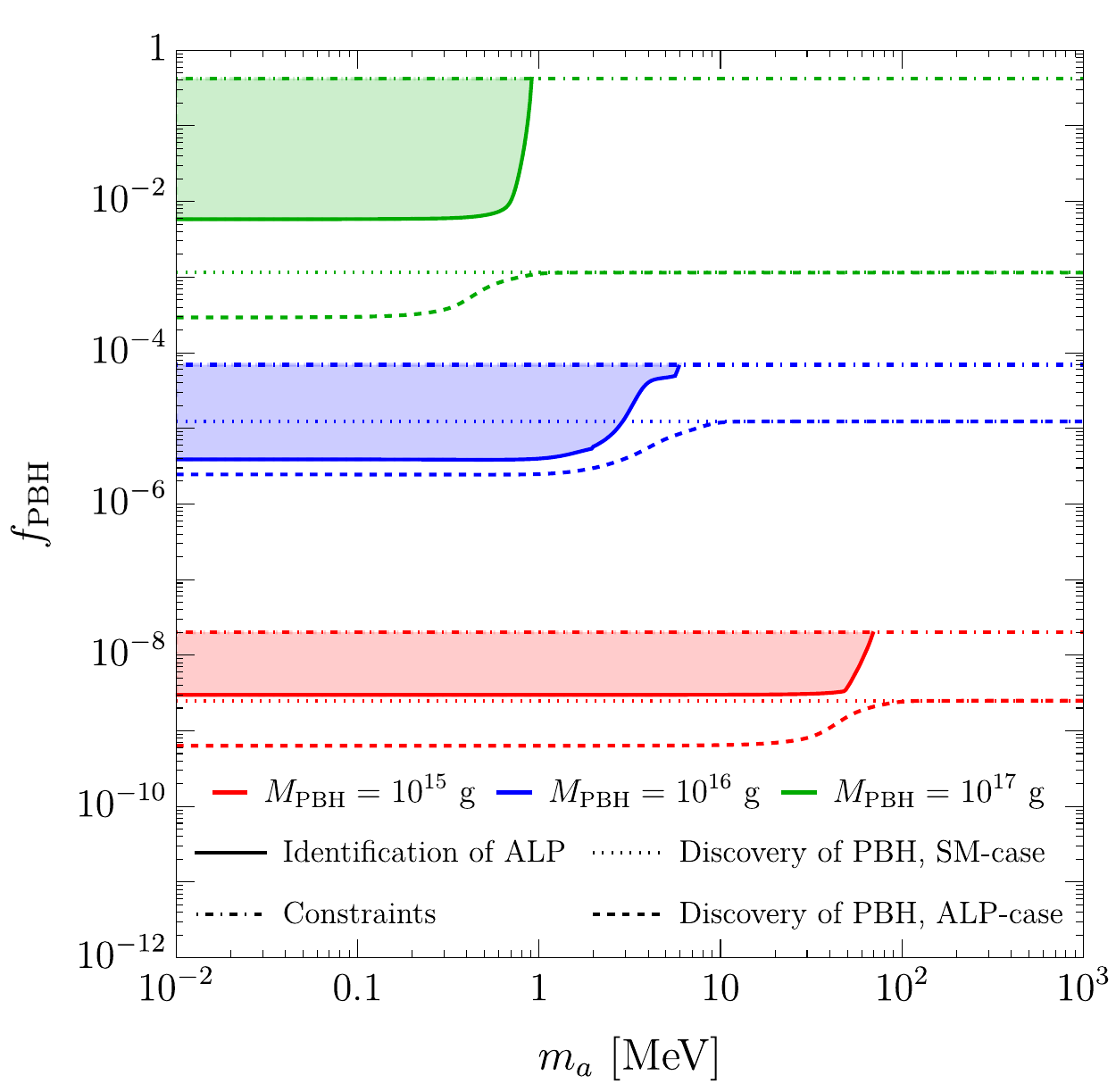}
\caption{PBH differentiability bounds in the $f_\mathrm{PBH}$ vs $M_\mathrm{PBH}$ ({\it left}) and $f_\mathrm{PBH}$ vs $m_a$ ({\it right}) planes. Various curves correspond to the capability to distinguish a PBH signal with the SM-case ({\it short-dash}) and the ALP-case ({\it long-dash}) from the background. Also included are the the ALP-case vs SM-case distinguishability ({\it solid}) and past experimental constraints ({\it left: shaded grey} and {\it right: dot-dashed}). In the $m_a$ plane, the shaded region highlights the region that can distinguish an ALP-case signal from an SM-case signal with the far right edge corresponding to the maximum $m_a$ allowed.}
\label{fig:PBHparamBounds}
\end{figure*}

In order to determine whether a particular model is observable (distinguishable from the astrophysical background), we perform the analysis described above using the background (no PBHs present) as the ``true" model. The parameter space of the test model is then scanned over in order to determine the parameter values such that the likelihood differs by a specified significance.\footnote{During the scan, we assume that the likelihood always decreases with increasing $f_\mathrm{PBH}$.} These results correspond to the PBH discovery bounds in Fig.~\ref{fig:PBHparamBounds} and will be discussed later. If the population of PBH is above the corresponding line, then the PBH will be bright enough that they are distinguishable from the background. Note that for a particular $M_\mathrm{PBH}$ and $m_a$, the ALP-case bound is always equal to or stronger than the SM-case bound. This is due to the PBHs producing the same SM particles; however, with the introduction of ALPs, there is an additional degree of freedom which results in a brighter, thus easier to observe, signal. Also note that for heavier PBHs or heavier ALPs, the ALP-case asymptotes to the SM-case while the other mass is held fixed. This is due to the ALP degree of freedom becoming exponentially suppressed when the Hawking temperature is much lower than $m_a$.

In addition to determining the point at which a given signal is observable, it is also convenient to discuss the parameter space where the ALP-case is distinguishable from the SM-case. To illustrate this, Fig.~\ref{fig:PBHparamContour} shows 3$\sigma$ contours from e-ASTROGAM constraining the PBH parameter space using Galactic Center data assuming that the underlying ``true" model is the ALP-case with $m_a=10^{-1}\;\mathrm{MeV}$, $M_\mathrm{PBH}=10^{16}\;\mathrm{g}$, and with various $f=f_\mathrm{PBH,true}$. The $y$-axis corresponds to the ratio between the test PBH fraction, $f_\mathrm{PBH,test}$, and $f_\mathrm{PBH,true}$.\footnote{The ratio was chosen for display purposes so that the true model is a fixed point in the figure rather than drifting. This has the additional benefit that no constraining curves intersect, but the disadvantage is that the upper boundaries (when there is insufficient signal to make any observation) become separate lines instead of a single result representing an upper bound on the parameter space.} As $f_\mathrm{PBH,true}$ increases, the signal-to-noise ratio also increases, thus strengthening the observability of the signal and the constraining capabilities of an experiment. In Fig.~\ref{fig:PBHparamContour}, the ``$\star$" indicates the location of the ``true" model. The two panels correspond to constraining the observation with the same model ({\it left}: ALP-case) and a different model ({\it right}: SM-case). The contours indicate the region of parameter space inside which can replicate the observational signal at the desired confidence. If the signal is too insufficient, only upper bounds can be placed. This is observed in the figure by the un-closed contours. As $f_\mathrm{PBH,true}$ increases, the signal increases and eventually lower bounds can be placed, as indicated by the closing of the contours. With further increases to $f_\mathrm{PBH,true}$, the contours shrink. If the test model matches the true model, the contours continuously shrink and eventually become infinitesimally small as the parameters of the true model are precisely determined ({\it left}). However, if the test model does not match, the contours will eventually collapse (no contours exist for $-\log_{10}(f_{\rm PBH,true})\leq5.40$) and no model parameter values will be able to replicate the signal ({\it right}). Note that these contours do not refer to those for constraining a particular PBH signal over the background, but rather the capability to confine PBH parameters given a particular observation. In order for a model to be distinguishable from another, the parameter space of one of the models must completely disappear. We utilize this behavior by assuming that there is a PBH signal (the chosen underlying model is the ALP-case) and scan over models where PBH only produce the SM-case. In the scan, we search for the value of $f_\mathrm{PBH,true}$ where the SM-case parameter space completely disappears. This is equivalent to searching for the parameter values where the likelihood of the bestfit model is equal to the significance criteria.\footnote{In this scan, it was assumed that the likelihood was single-peaked in both $M_\mathrm{PBH}$ and $f_\mathrm{PBH}$. As observed in \Fig{fig:PBHparamContour}, this may not be the case; however, tests indicate that it would usually only lead to minor fluctuations in the final result. The oscillating behavior observed in the low signal-to-noise regions of \Fig{fig:PBHparamContour} ultimately produces these islands as the parameter space is squeezed as it is constrained. This wave-like behavior appears to be a result of the photon energy binning and can be reduced by increased experimental resolution to allow for finer bins.}

We are therefore able to define the $f_\mathrm{PBH}$ for a given $M_\mathrm{PBH}$ and $m_a$ where the ALP-case is discernable from SM-case. This value of $f_\mathrm{PBH}$ is the black hole fraction needed for the allowable SM-case PBH parameter space to completely disappear at a given confidence interval. This discernability threshold is shown in \Fig{fig:PBHparamBounds} for various $M_\mathrm{PBH}$ ({\it left}) and $m_a$ ({\it right}) with the other parameter fixed.

In the $M_\mathrm{PBH}$ plane of \Fig{fig:PBHparamBounds}, all parameter values above the solid curves correspond to the region where a ALP-case signal can be distinguished from the SM-case. Also included are the corresponding lower limits for $f_\mathrm{PBH}$ in order to distinguish a PBH signal from the background for the ALP-case ({\it long dash}) and the SM-case ({\it short dash}). As expected, the ALP-case vs SM-case distinguishability line is always higher than the ALP-case identification line as it must first be observable before more information can be extracted. In addition, the ALP-case identification line is always beneath the SM-case identification line. This is because it is a brighter signal as mentioned previously. Other features of note are that the ALP-case and the SM-case identification lines merge for large $M_\mathrm{PBH}$; this is because the ALPs become exponentially suppressed as the Hawking temperature drops below their mass. This transition is also one of the causes for the distinguishability lines rapidly losing sensitivity as the ALP-case and the SM-case signals become nearly identical. The other cause is related to the experimental sensitivity as signal either leaves the detector's range (such as with high PBH masses) or enters regions of low sensitivity (the convergence of discovery lines near $5 \times 10^{15}\;\mathrm{g}$ are a result of the extra ALP decay photons residing in a region of low detector sensitivity). Also included are current PBH bounds for the SM-case from \cite{Clark:2018ghm,  Coogan:2020tuf}, and see \cite{Boudaud:2018hqb, Poulin:2016anj,Clark:2016nst, DeRocco:2019fjq, Laha:2019ssq, Carr:2009jm, Kim:2020ngi, Saha:2021pqf, Laha:2020ivk, Green:2020jor} for other constraints in the  asteroid-mass window.\footnote{While these bounds will change with the introduction of the ALP, the alteration is expected not to be larger than a factor of 2 (corresponding to the massless case) due to the ALPs simply adding an additional degree of freedom.}

In the $m_a$ plane of \Fig{fig:PBHparamBounds}, each line style corresponds to the same result as in the $f_\mathrm{PBH}$ panel. Both the ALP-case vs SM-case distinguishability curves ({\it solid}) and the ALP-case PBH background discovery curves ({\it long dash}) are flat for small $m_a$ as they are essentially massless when compared with their energies. For large masses, the reverse is true and they are both nonrelativistic and exponentially suppressed as the Hawking temperature drops below their mass. This leads to another flat region for the identification from background curve as it plateaus and becomes identical to the SM-case identification curve ({\it short dash}). On the other hand, the ALP-case vs SM-case distinguishability curve rapidly increases due to the lack of distinguishing features. In addition, current SM-case PBH constraints are also plotted ({\it dot-dashed}), above which SM-case PBHs have already been constrained. The shaded regions correspond to the allowed parameter space for distinguishing the ALP-case from the SM-case, and the rightmost edge is the high $m_a$ bound of the regions shown in Fig.~\ref{fig:ALPspectrum}.

\section{Conclusions and Outlook}\label{sec.conclusions}

PBHs are a plausible candidate for a fraction or all of the observed DM density. They are also a unique source for producing BSM particles, even if the new particles have a tiny coupling to the SM sector. In this work, we discuss the possibility of observing Galactic Center gamma-ray signals both from direct PBH radiation and from the decay of ALPs produced by PBHs. Under the assumption of monochromatic PBH mass, we show that with $M_{\rm PBH}=10^{15-17}$~g, and $m_a\lsim 60$~MeV, e-ASTROGAM has a chance to observe both types of gamma-ray signals and distinguish the total signal from PBHs with arbitrary choices of the $M_{\rm PBH}$ and abundance $f_{\rm PBH}$. Our findings show that future detectors such as the e-ASTROGAM and AMEGO experiments can explore both PBH and axion physics, even for ALPs that satisfies all existing constraints and have no ambient presence in the universe today. One can also consider the use of the extragalactic gamma-ray background when performing this analysis. Before considering differences in the PBH distribution and the astrophysical background, this has the potential of increasing the coverage of smaller $g_{a\gamma\gamma}$ by up to three orders of magnitude due to increasing the allowable ALP decay length to be the Horizon size. We leave this for future work.

\begin{figure}[ht]
\centering
\includegraphics[width=0.98\columnwidth]{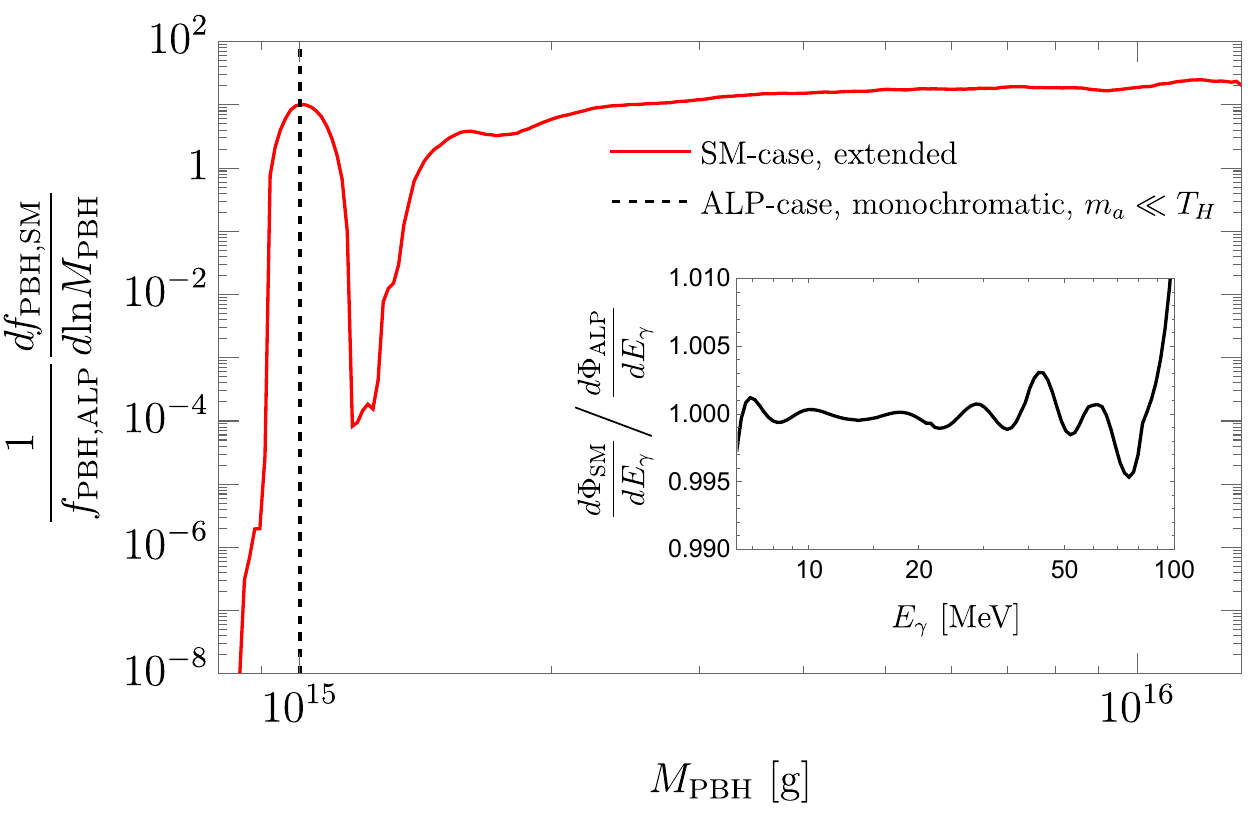}
\caption{The extended black hole mass distribution that mimics the photon spectrum for the ALP-case with $m_a\ll T_H$, $\MBH=10^{15}$ g, and $6 ~\MeV<E_\gamma<100~\MeV$. The peak near the $\MBH=10^{15}$ g explains the primary photon peak, while the plateau at large masses is for the peak from ALP decays. To show the goodness of the fit, we show the the ratio of the gamma-ray flux between SM-case with the extended mass function and the ALP-case in the plot inside. As a result, $\fPBH$ for the SM-case is more than 200 times larger than $\fPBH$ for the ALP-case.}
\label{fig:BHmassfunction}
\end{figure}

To make a stronger statement in distinguishing the ALP-case signal from the SM-case, we need to perform the analysis with a more generalized PBH mass function. In this case, a non-trivial mass function may produce the ``double peak" feature as in the ALP-case signal (See \Fig{fig:ALPspectrum}.), making it difficult to confirm the existence of the ALP decay. However, mimicking the ALP signal with an extended PBH mass function may not be trivial. Since the ALP decay mainly contributes to the softer gamma-ray spectrum as in Fig.~\ref{fig:ALPspectrum}, it requires a large density of massive PBHs to fully reproduce the lower energy tail of the ALP signal. In \Fig{fig:BHmassfunction}, we show an example of mimicking the ALP-case signal with $M_{\rm PBH}=10^{15}$~g and massless ALPs by an extended PBH mass function in the SM-case. When trying to reproduce the original gamma-ray spectrum down to $E_\gamma=6$~MeV, the plateau of the mass spectrum (red) in the higher PBH mass makes the $f_{\rm PBH}$ about $200$ times larger than in the ALP-case. It can also be hard to realize the exotic feature of the mass spectrum that has a dip right next to the original PBH mass within the context of a cosmological model.

Even if the mass function in the SM-case mimics the ALP-case signal and satisfies the DM bound, we may use the gravitational wave (GW) signal to distinguish the ALP signal from the SM radiation. If the PBHs are produced by large primordial curvature perturbations, these will differ between the ALP-case and the SM-case which give rise to the same gamma-ray signal, since the corresponding PBH spectra are
different. As studied in Ref.~\cite{Agashe:2022jgk}, large curvature perturbations that 
are associated with the production of the PBHs 
with the visible gamma-ray signal at e-ASTROGAM will source GW signals well above the sensitivity of future detectors. Combining the above two facts, the GW signals for the two cases will then be different, even though the gamma-ray signal is the same (the latter by design).
In other words, there is a chance to distinguish the two scenarios by correlating the gamma-ray and GW signals. We leave these studies for future directions.
\\
\\
{\it Note added}: As we were finalizing the paper, we came to know of work by Y.~Jho, T.~G.~Kim, J-C.~Park, S-C.~Park and Y.~Park on a similar topic \cite{Jho:2022wxd}.

\section*{\uppercase{Acknowledgements}}
We would like to thank Andrea Caputo, Henrike Fleischhack, Subhajit Ghosh, Shmuel Nussinov, Nadav Outmezguine, and Edoardo Vitagliano for useful discussions and Jong-Chul Park {\it et al.} for informing us about their work on a similar topic. We are also grateful to Debasish Borah, Nayan Das, and Prantik Sarmah for pointing out an error in \Fig{fig:ALPparam}. KA and JHC were supported in part by the NSF grant PHY-2210361 and by the Maryland Center for Fundamental Physics. JHC is also supported in part by JHU Joint Postdoc Fund. BD is supported in part by DOE grant DE-SC0010813. TX is supported in part by DOE Grant desc0009956 and  by the Israel Science Foundation (grant No. 1112/17). YT is supported by the NSF grant PHY-2112540.

\bibliographystyle{utphys}
\bibliography{bibliography.bib}

\providecommand{\href}[2]{#2}\begingroup\raggedright\begin{thebibliography}{100}

\bibitem{e-ASTROGAM:2016bph}
{\bfseries e-ASTROGAM} Collaboration, A.~De~Angelis {\em et~al.}, ``{The
  e-ASTROGAM mission},''
  \href{http://dx.doi.org/10.1007/s10686-017-9533-6}{{\em Exper. Astron.}
  {\bfseries 44} no.~1, (2017) 25--82},
  \href{http://arxiv.org/abs/1611.02232}{{\ttfamily arXiv:1611.02232
  [astro-ph.HE]}}.

\bibitem{AMEGO:2019gny}
{\bfseries AMEGO} Collaboration, R.~Caputo {\em et~al.}, ``{All-sky Medium
  Energy Gamma-ray Observatory: Exploring the Extreme Multimessenger
  Universe},'' \href{http://arxiv.org/abs/1907.07558}{{\ttfamily
  arXiv:1907.07558 [astro-ph.IM]}}.

\bibitem{Boehm:2002yz}
C.~Boehm, T.~A. Ensslin, and J.~Silk, ``{Can Annihilating dark matter be
  lighter than a few GeVs?},''
  \href{http://dx.doi.org/10.1088/0954-3899/30/3/004}{{\em J. Phys. G}
  {\bfseries 30} (2004) 279--286},
  \href{http://arxiv.org/abs/astro-ph/0208458}{{\ttfamily
  arXiv:astro-ph/0208458}}.

\bibitem{Beacom:2004pe}
J.~F. Beacom, N.~F. Bell, and G.~Bertone, ``{Gamma-ray constraint on Galactic
  positron production by MeV dark matter},''
  \href{http://dx.doi.org/10.1103/PhysRevLett.94.171301}{{\em Phys. Rev. Lett.}
  {\bfseries 94} (2005) 171301},
  \href{http://arxiv.org/abs/astro-ph/0409403}{{\ttfamily
  arXiv:astro-ph/0409403}}.

\bibitem{Finkbeiner:2007kk}
D.~P. Finkbeiner and N.~Weiner, ``{Exciting Dark Matter and the INTEGRAL/SPI
  511 keV signal},'' \href{http://dx.doi.org/10.1103/PhysRevD.76.083519}{{\em
  Phys. Rev. D} {\bfseries 76} (2007) 083519},
  \href{http://arxiv.org/abs/astro-ph/0702587}{{\ttfamily
  arXiv:astro-ph/0702587}}.

\bibitem{Essig:2009jx}
R.~Essig, N.~Sehgal, and L.~E. Strigari, ``{Bounds on Cross-sections and
  Lifetimes for Dark Matter Annihilation and Decay into Charged Leptons from
  Gamma-ray Observations of Dwarf Galaxies},''
  \href{http://dx.doi.org/10.1103/PhysRevD.80.023506}{{\em Phys. Rev. D}
  {\bfseries 80} (2009) 023506},
  \href{http://arxiv.org/abs/0902.4750}{{\ttfamily arXiv:0902.4750 [hep-ph]}}.

\bibitem{Essig:2013goa}
R.~Essig, E.~Kuflik, S.~D. McDermott, T.~Volansky, and K.~M. Zurek,
  ``{Constraining Light Dark Matter with Diffuse X-Ray and Gamma-Ray
  Observations},'' \href{http://dx.doi.org/10.1007/JHEP11(2013)193}{{\em JHEP}
  {\bfseries 11} (2013) 193}, \href{http://arxiv.org/abs/1309.4091}{{\ttfamily
  arXiv:1309.4091 [hep-ph]}}.

\bibitem{Boddy:2015efa}
K.~K. Boddy and J.~Kumar, ``{Indirect Detection of Dark Matter Using MeV-Range
  Gamma-Ray Telescopes},''
  \href{http://dx.doi.org/10.1103/PhysRevD.92.023533}{{\em Phys. Rev. D}
  {\bfseries 92} no.~2, (2015) 023533},
  \href{http://arxiv.org/abs/1504.04024}{{\ttfamily arXiv:1504.04024
  [astro-ph.CO]}}.

\bibitem{Laha:2020ivk}
R.~Laha, J.~B. Mu\~noz, and T.~R. Slatyer, ``{INTEGRAL constraints on
  primordial black holes and particle dark matter},''
  \href{http://dx.doi.org/10.1103/PhysRevD.101.123514}{{\em Phys. Rev. D}
  {\bfseries 101} no.~12, (2020) 123514},
  \href{http://arxiv.org/abs/2004.00627}{{\ttfamily arXiv:2004.00627
  [astro-ph.CO]}}.

\bibitem{Carr:2009jm}
B.~J. Carr, K.~Kohri, Y.~Sendouda, and J.~Yokoyama, ``{New cosmological
  constraints on primordial black holes},''
  \href{http://dx.doi.org/10.1103/PhysRevD.81.104019}{{\em Phys. Rev. D}
  {\bfseries 81} (2010) 104019},
  \href{http://arxiv.org/abs/0912.5297}{{\ttfamily arXiv:0912.5297
  [astro-ph.CO]}}.

\bibitem{Laha:2019ssq}
R.~Laha, ``{Primordial Black Holes as a Dark Matter Candidate Are Severely
  Constrained by the Galactic Center 511 keV $\gamma$ -Ray Line},''
  \href{http://dx.doi.org/10.1103/PhysRevLett.123.251101}{{\em Phys. Rev.
  Lett.} {\bfseries 123} no.~25, (2019) 251101},
  \href{http://arxiv.org/abs/1906.09994}{{\ttfamily arXiv:1906.09994
  [astro-ph.HE]}}.

\bibitem{Carr:2020gox}
B.~Carr, K.~Kohri, Y.~Sendouda, and J.~Yokoyama, ``{Constraints on primordial
  black holes},'' \href{http://dx.doi.org/10.1088/1361-6633/ac1e31}{{\em Rept.
  Prog. Phys.} {\bfseries 84} no.~11, (2021) 116902},
  \href{http://arxiv.org/abs/2002.12778}{{\ttfamily arXiv:2002.12778
  [astro-ph.CO]}}.

\bibitem{Coogan:2020tuf}
A.~Coogan, L.~Morrison, and S.~Profumo, ``{Direct Detection of Hawking
  Radiation from Asteroid-Mass Primordial Black Holes},''
  \href{http://dx.doi.org/10.1103/PhysRevLett.126.171101}{{\em Phys. Rev.
  Lett.} {\bfseries 126} no.~17, (2021) 171101},
  \href{http://arxiv.org/abs/2010.04797}{{\ttfamily arXiv:2010.04797
  [astro-ph.CO]}}.

\bibitem{Ray:2021mxu}
A.~Ray, R.~Laha, J.~B. Mu\~noz, and R.~Caputo, ``{Near future MeV telescopes
  can discover asteroid-mass primordial black hole dark matter},''
  \href{http://dx.doi.org/10.1103/PhysRevD.104.023516}{{\em Phys. Rev. D}
  {\bfseries 104} no.~2, (2021) 023516},
  \href{http://arxiv.org/abs/2102.06714}{{\ttfamily arXiv:2102.06714
  [astro-ph.CO]}}.

\bibitem{Keith:2022sow}
C.~Keith, D.~Hooper, T.~Linden, and R.~Liu, ``{Sensitivity of future gamma-ray
  telescopes to primordial black holes},''
  \href{http://dx.doi.org/10.1103/PhysRevD.106.043003}{{\em Phys. Rev. D}
  {\bfseries 106} no.~4, (2022) 043003},
  \href{http://arxiv.org/abs/2204.05337}{{\ttfamily arXiv:2204.05337
  [astro-ph.HE]}}.

\bibitem{Tan:2022lbm}
X.-H. Tan, Y.-J. Yan, T.~Qiu, and J.-Q. Xia, ``{Searching for the Signal of a
  Primordial Black Hole from CMB Lensing and \ensuremath{\gamma}-Ray
  Emissions},'' \href{http://dx.doi.org/10.3847/2041-8213/ac9668}{{\em
  Astrophys. J. Lett.} {\bfseries 939} no.~1, (2022) L15},
  \href{http://arxiv.org/abs/2209.15222}{{\ttfamily arXiv:2209.15222
  [astro-ph.CO]}}.

\bibitem{Caputo:2022dkz}
A.~Caputo, M.~Negro, M.~Regis, and M.~Taoso, ``{Dark Matter prospects with
  COSI: ALPs, PBHs and sub-GeV Dark Matter},''
  \href{http://arxiv.org/abs/2210.09310}{{\ttfamily arXiv:2210.09310
  [hep-ph]}}.

\bibitem{Carr:2016drx}
B.~Carr, F.~Kuhnel, and M.~Sandstad, ``{Primordial Black Holes as Dark
  Matter},'' \href{http://dx.doi.org/10.1103/PhysRevD.94.083504}{{\em Phys.
  Rev. D} {\bfseries 94} no.~8, (2016) 083504},
  \href{http://arxiv.org/abs/1607.06077}{{\ttfamily arXiv:1607.06077
  [astro-ph.CO]}}.

\bibitem{Carr:1975qj}
B.~J. Carr, ``{The Primordial black hole mass spectrum},''
  \href{http://dx.doi.org/10.1086/153853}{{\em Astrophys. J.} {\bfseries 201}
  (1975) 1--19}.

\bibitem{Ivanov:1994pa}
P.~Ivanov, P.~Naselsky, and I.~Novikov, ``{Inflation and primordial black holes
  as dark matter},'' \href{http://dx.doi.org/10.1103/PhysRevD.50.7173}{{\em
  Phys. Rev. D} {\bfseries 50} (1994) 7173--7178}.

\bibitem{Garcia-Bellido:1996mdl}
J.~Garcia-Bellido, A.~D. Linde, and D.~Wands, ``{Density perturbations and
  black hole formation in hybrid inflation},''
  \href{http://dx.doi.org/10.1103/PhysRevD.54.6040}{{\em Phys. Rev. D}
  {\bfseries 54} (1996) 6040--6058},
  \href{http://arxiv.org/abs/astro-ph/9605094}{{\ttfamily
  arXiv:astro-ph/9605094}}.

\bibitem{Silk:1986vc}
J.~Silk and M.~S. Turner, ``{Double Inflation},''
  \href{http://dx.doi.org/10.1103/PhysRevD.35.419}{{\em Phys. Rev. D}
  {\bfseries 35} (1987) 419}.

\bibitem{Kawasaki:1997ju}
M.~Kawasaki, N.~Sugiyama, and T.~Yanagida, ``{Primordial black hole formation
  in a double inflation model in supergravity},''
  \href{http://dx.doi.org/10.1103/PhysRevD.57.6050}{{\em Phys. Rev. D}
  {\bfseries 57} (1998) 6050--6056},
  \href{http://arxiv.org/abs/hep-ph/9710259}{{\ttfamily arXiv:hep-ph/9710259}}.

\bibitem{Yokoyama:1995ex}
J.~Yokoyama, ``{Formation of MACHO primordial black holes in inflationary
  cosmology},'' {\em Astron. Astrophys.} {\bfseries 318} (1997) 673,
  \href{http://arxiv.org/abs/astro-ph/9509027}{{\ttfamily
  arXiv:astro-ph/9509027}}.

\bibitem{Pi:2017gih}
S.~Pi, Y.-l. Zhang, Q.-G. Huang, and M.~Sasaki, ``{Scalaron from $R^2$-gravity
  as a heavy field},''
  \href{http://dx.doi.org/10.1088/1475-7516/2018/05/042}{{\em JCAP} {\bfseries
  05} (2018) 042}, \href{http://arxiv.org/abs/1712.09896}{{\ttfamily
  arXiv:1712.09896 [astro-ph.CO]}}.

\bibitem{Hertzberg:2017dkh}
M.~P. Hertzberg and M.~Yamada, ``{Primordial Black Holes from Polynomial
  Potentials in Single Field Inflation},''
  \href{http://dx.doi.org/10.1103/PhysRevD.97.083509}{{\em Phys. Rev. D}
  {\bfseries 97} no.~8, (2018) 083509},
  \href{http://arxiv.org/abs/1712.09750}{{\ttfamily arXiv:1712.09750
  [astro-ph.CO]}}.

\bibitem{Ozsoy:2018flq}
O.~\"Ozsoy, S.~Parameswaran, G.~Tasinato, and I.~Zavala, ``{Mechanisms for
  Primordial Black Hole Production in String Theory},''
  \href{http://dx.doi.org/10.1088/1475-7516/2018/07/005}{{\em JCAP} {\bfseries
  07} (2018) 005}, \href{http://arxiv.org/abs/1803.07626}{{\ttfamily
  arXiv:1803.07626 [hep-th]}}.

\bibitem{Cicoli:2018asa}
M.~Cicoli, V.~A. Diaz, and F.~G. Pedro, ``{Primordial Black Holes from String
  Inflation},'' \href{http://dx.doi.org/10.1088/1475-7516/2018/06/034}{{\em
  JCAP} {\bfseries 06} (2018) 034},
  \href{http://arxiv.org/abs/1803.02837}{{\ttfamily arXiv:1803.02837
  [hep-th]}}.

\bibitem{Ashoorioon:2019xqc}
A.~Ashoorioon, A.~Rostami, and J.~T. Firouzjaee, ``{EFT compatible PBHs:
  effective spawning of the seeds for primordial black holes during
  inflation},'' \href{http://dx.doi.org/10.1007/JHEP07(2021)087}{{\em JHEP}
  {\bfseries 07} (2021) 087}, \href{http://arxiv.org/abs/1912.13326}{{\ttfamily
  arXiv:1912.13326 [astro-ph.CO]}}.

\bibitem{Hawking:1982ga}
S.~W. Hawking, I.~G. Moss, and J.~M. Stewart, ``{Bubble Collisions in the Very
  Early Universe},'' \href{http://dx.doi.org/10.1103/PhysRevD.26.2681}{{\em
  Phys. Rev. D} {\bfseries 26} (1982) 2681}.

\bibitem{Crawford:1982yz}
M.~Crawford and D.~N. Schramm, ``{Spontaneous Generation of Density
  Perturbations in the Early Universe},''
  \href{http://dx.doi.org/10.1038/298538a0}{{\em Nature} {\bfseries 298} (1982)
  538--540}.

\bibitem{Kodama:1982sf}
H.~Kodama, M.~Sasaki, and K.~Sato, ``{Abundance of Primordial Holes Produced by
  Cosmological First Order Phase Transition},''
  \href{http://dx.doi.org/10.1143/PTP.68.1979}{{\em Prog. Theor. Phys.}
  {\bfseries 68} (1982) 1979}.

\bibitem{Moss:1994pi}
I.~G. Moss, ``{Black hole formation from colliding bubbles},''
  \href{http://arxiv.org/abs/gr-qc/9405045}{{\ttfamily arXiv:gr-qc/9405045}}.

\bibitem{Freivogel:2007fx}
B.~Freivogel, G.~T. Horowitz, and S.~Shenker, ``{Colliding with a crunching
  bubble},'' \href{http://dx.doi.org/10.1088/1126-6708/2007/05/090}{{\em JHEP}
  {\bfseries 05} (2007) 090},
  \href{http://arxiv.org/abs/hep-th/0703146}{{\ttfamily arXiv:hep-th/0703146}}.

\bibitem{Johnson:2011wt}
M.~C. Johnson, H.~V. Peiris, and L.~Lehner, ``{Determining the outcome of
  cosmic bubble collisions in full General Relativity},''
  \href{http://dx.doi.org/10.1103/PhysRevD.85.083516}{{\em Phys. Rev. D}
  {\bfseries 85} (2012) 083516},
  \href{http://arxiv.org/abs/1112.4487}{{\ttfamily arXiv:1112.4487 [hep-th]}}.

\bibitem{Kusenko:2020pcg}
A.~Kusenko, M.~Sasaki, S.~Sugiyama, M.~Takada, V.~Takhistov, and E.~Vitagliano,
  ``{Exploring Primordial Black Holes from the Multiverse with Optical
  Telescopes},'' \href{http://dx.doi.org/10.1103/PhysRevLett.125.181304}{{\em
  Phys. Rev. Lett.} {\bfseries 125} (2020) 181304},
  \href{http://arxiv.org/abs/2001.09160}{{\ttfamily arXiv:2001.09160
  [astro-ph.CO]}}.

\bibitem{Ashoorioon:2020hln}
A.~Ashoorioon, A.~Rostami, and J.~T. Firouzjaee, ``{Examining the end of
  inflation with primordial black holes mass distribution and gravitational
  waves},'' \href{http://dx.doi.org/10.1103/PhysRevD.103.123512}{{\em Phys.
  Rev. D} {\bfseries 103} (2021) 123512},
  \href{http://arxiv.org/abs/2012.02817}{{\ttfamily arXiv:2012.02817
  [astro-ph.CO]}}.

\bibitem{Baker:2021nyl}
M.~J. Baker, M.~Breitbach, J.~Kopp, and L.~Mittnacht, ``{Primordial Black Holes
  from First-Order Cosmological Phase Transitions},''
  \href{http://arxiv.org/abs/2105.07481}{{\ttfamily arXiv:2105.07481
  [astro-ph.CO]}}.

\bibitem{Baker:2021sno}
M.~J. Baker, M.~Breitbach, J.~Kopp, and L.~Mittnacht, ``{Detailed Calculation
  of Primordial Black Hole Formation During First-Order Cosmological Phase
  Transitions},'' \href{http://arxiv.org/abs/2110.00005}{{\ttfamily
  arXiv:2110.00005 [astro-ph.CO]}}.

\bibitem{Kawana:2021tde}
K.~Kawana and K.-P. Xie, ``{Primordial black holes from a cosmic phase
  transition: The collapse of Fermi-balls},''
  \href{http://dx.doi.org/10.1016/j.physletb.2021.136791}{{\em Phys. Lett. B}
  {\bfseries 824} (2022) 136791},
  \href{http://arxiv.org/abs/2106.00111}{{\ttfamily arXiv:2106.00111
  [astro-ph.CO]}}.

\bibitem{Huang:2022him}
P.~Huang and K.-P. Xie, ``{Primordial black holes from an electroweak phase
  transition},'' \href{http://dx.doi.org/10.1103/PhysRevD.105.115033}{{\em
  Phys. Rev. D} {\bfseries 105} no.~11, (2022) 115033},
  \href{http://arxiv.org/abs/2201.07243}{{\ttfamily arXiv:2201.07243
  [hep-ph]}}.

\bibitem{Lu:2022paj}
P.~Lu, K.~Kawana, and K.-P. Xie, ``{Old phase remnants in first-order phase
  transitions},'' \href{http://dx.doi.org/10.1103/PhysRevD.105.123503}{{\em
  Phys. Rev. D} {\bfseries 105} no.~12, (2022) 123503},
  \href{http://arxiv.org/abs/2202.03439}{{\ttfamily arXiv:2202.03439
  [astro-ph.CO]}}.

\bibitem{Ashoorioon:2022raz}
A.~Ashoorioon, K.~Rezazadeh, and A.~Rostami, ``{NANOGrav signal from the end of
  inflation and the LIGO mass and heavier primordial black holes},''
  \href{http://dx.doi.org/10.1016/j.physletb.2022.137542}{{\em Phys. Lett. B}
  {\bfseries 835} (2022) 137542},
  \href{http://arxiv.org/abs/2202.01131}{{\ttfamily arXiv:2202.01131
  [astro-ph.CO]}}.

\bibitem{Cotner:2016cvr}
E.~Cotner and A.~Kusenko, ``{Primordial black holes from supersymmetry in the
  early universe},''
  \href{http://dx.doi.org/10.1103/PhysRevLett.119.031103}{{\em Phys. Rev.
  Lett.} {\bfseries 119} no.~3, (2017) 031103},
  \href{http://arxiv.org/abs/1612.02529}{{\ttfamily arXiv:1612.02529
  [astro-ph.CO]}}.

\bibitem{Cotner:2017tir}
E.~Cotner and A.~Kusenko, ``{Primordial black holes from scalar field evolution
  in the early universe},''
  \href{http://dx.doi.org/10.1103/PhysRevD.96.103002}{{\em Phys. Rev. D}
  {\bfseries 96} no.~10, (2017) 103002},
  \href{http://arxiv.org/abs/1706.09003}{{\ttfamily arXiv:1706.09003
  [astro-ph.CO]}}.

\bibitem{Cotner:2018vug}
E.~Cotner, A.~Kusenko, and V.~Takhistov, ``{Primordial Black Holes from
  Inflaton Fragmentation into Oscillons},''
  \href{http://dx.doi.org/10.1103/PhysRevD.98.083513}{{\em Phys. Rev. D}
  {\bfseries 98} no.~8, (2018) 083513},
  \href{http://arxiv.org/abs/1801.03321}{{\ttfamily arXiv:1801.03321
  [astro-ph.CO]}}.

\bibitem{Cotner:2019ykd}
E.~Cotner, A.~Kusenko, M.~Sasaki, and V.~Takhistov, ``{Analytic Description of
  Primordial Black Hole Formation from Scalar Field Fragmentation},''
  \href{http://dx.doi.org/10.1088/1475-7516/2019/10/077}{{\em JCAP} {\bfseries
  10} (2019) 077}, \href{http://arxiv.org/abs/1907.10613}{{\ttfamily
  arXiv:1907.10613 [astro-ph.CO]}}.

\bibitem{Hawking:1987bn}
S.~W. Hawking, ``{Black Holes From Cosmic Strings},''
  \href{http://dx.doi.org/10.1016/0370-2693(89)90206-2}{{\em Phys. Lett. B}
  {\bfseries 231} (1989) 237--239}.

\bibitem{Polnarev:1988dh}
A.~Polnarev and R.~Zembowicz, ``{Formation of Primordial Black Holes by Cosmic
  Strings},'' \href{http://dx.doi.org/10.1103/PhysRevD.43.1106}{{\em Phys. Rev.
  D} {\bfseries 43} (1991) 1106--1109}.

\bibitem{MacGibbon:1997pu}
J.~H. MacGibbon, R.~H. Brandenberger, and U.~F. Wichoski, ``{Limits on black
  hole formation from cosmic string loops},''
  \href{http://dx.doi.org/10.1103/PhysRevD.57.2158}{{\em Phys. Rev. D}
  {\bfseries 57} (1998) 2158--2165},
  \href{http://arxiv.org/abs/astro-ph/9707146}{{\ttfamily
  arXiv:astro-ph/9707146}}.

\bibitem{Brandenberger:2021zvn}
R.~Brandenberger, B.~Cyr, and H.~Jiao, ``{Intermediate mass black hole seeds
  from cosmic string loops},''
  \href{http://dx.doi.org/10.1103/PhysRevD.104.123501}{{\em Phys. Rev. D}
  {\bfseries 104} no.~12, (2021) 123501},
  \href{http://arxiv.org/abs/2103.14057}{{\ttfamily arXiv:2103.14057
  [astro-ph.CO]}}.

\bibitem{Rubin:2000dq}
S.~G. Rubin, M.~Y. Khlopov, and A.~S. Sakharov, ``{Primordial black holes from
  nonequilibrium second order phase transition},'' {\em Grav. Cosmol.}
  {\bfseries 6} (2000) 51--58,
  \href{http://arxiv.org/abs/hep-ph/0005271}{{\ttfamily arXiv:hep-ph/0005271}}.

\bibitem{Rubin:2001yw}
S.~G. Rubin, A.~S. Sakharov, and M.~Y. Khlopov, ``{The Formation of primary
  galactic nuclei during phase transitions in the early universe},''
  \href{http://dx.doi.org/10.1134/1.1385631}{{\em J. Exp. Theor. Phys.}
  {\bfseries 91} (2001) 921--929},
  \href{http://arxiv.org/abs/hep-ph/0106187}{{\ttfamily arXiv:hep-ph/0106187}}.

\bibitem{Banks:2018ypk}
T.~Banks and W.~Fischler, ``{The holographic spacetime model of cosmology},''
  \href{http://dx.doi.org/10.1142/S0218271818460057}{{\em Int. J. Mod. Phys. D}
  {\bfseries 27} no.~14, (2018) 1846005},
  \href{http://arxiv.org/abs/1806.01749}{{\ttfamily arXiv:1806.01749
  [hep-th]}}.

\bibitem{Banks:2020dgx}
T.~Banks and W.~Fischler, ``{Primordial Black Holes as Dark Matter},''
  \href{http://arxiv.org/abs/2008.00327}{{\ttfamily arXiv:2008.00327
  [hep-th]}}.

\bibitem{Banks:2021lai}
T.~Banks and W.~Fischler, ``{Entropy and Black Holes in the Very Early
  Universe},'' \href{http://arxiv.org/abs/2109.05571}{{\ttfamily
  arXiv:2109.05571 [hep-th]}}.

\bibitem{Bell:1998jk}
N.~F. Bell and R.~R. Volkas, ``{Mirror matter and primordial black holes},''
  \href{http://dx.doi.org/10.1103/PhysRevD.59.107301}{{\em Phys. Rev. D}
  {\bfseries 59} (1999) 107301},
  \href{http://arxiv.org/abs/astro-ph/9812301}{{\ttfamily
  arXiv:astro-ph/9812301}}.

\bibitem{Allahverdi:2017sks}
R.~Allahverdi, J.~Dent, and J.~Osinski, ``{Nonthermal production of dark matter
  from primordial black holes},''
  \href{http://dx.doi.org/10.1103/PhysRevD.97.055013}{{\em Phys. Rev. D}
  {\bfseries 97} no.~5, (2018) 055013},
  \href{http://arxiv.org/abs/1711.10511}{{\ttfamily arXiv:1711.10511
  [astro-ph.CO]}}.

\bibitem{Lennon:2017tqq}
O.~Lennon, J.~March-Russell, R.~Petrossian-Byrne, and H.~Tillim, ``{Black Hole
  Genesis of Dark Matter},''
  \href{http://dx.doi.org/10.1088/1475-7516/2018/04/009}{{\em JCAP} {\bfseries
  04} (2018) 009}, \href{http://arxiv.org/abs/1712.07664}{{\ttfamily
  arXiv:1712.07664 [hep-ph]}}.

\bibitem{Hooper:2019gtx}
D.~Hooper, G.~Krnjaic, and S.~D. McDermott, ``{Dark Radiation and Superheavy
  Dark Matter from Black Hole Domination},''
  \href{http://dx.doi.org/10.1007/JHEP08(2019)001}{{\em JHEP} {\bfseries 08}
  (2019) 001}, \href{http://arxiv.org/abs/1905.01301}{{\ttfamily
  arXiv:1905.01301 [hep-ph]}}.

\bibitem{Gondolo:2020uqv}
P.~Gondolo, P.~Sandick, and B.~Shams Es~Haghi, ``{Effects of primordial black
  holes on dark matter models},''
  \href{http://dx.doi.org/10.1103/PhysRevD.102.095018}{{\em Phys. Rev. D}
  {\bfseries 102} no.~9, (2020) 095018},
  \href{http://arxiv.org/abs/2009.02424}{{\ttfamily arXiv:2009.02424
  [hep-ph]}}.

\bibitem{Cheek:2021odj}
A.~Cheek, L.~Heurtier, Y.~F. Perez-Gonzalez, and J.~Turner, ``{Primordial black
  hole evaporation and dark matter production. I. Solely Hawking radiation},''
  \href{http://dx.doi.org/10.1103/PhysRevD.105.015022}{{\em Phys. Rev. D}
  {\bfseries 105} no.~1, (2022) 015022},
  \href{http://arxiv.org/abs/2107.00013}{{\ttfamily arXiv:2107.00013
  [hep-ph]}}.

\bibitem{Cheek:2021cfe}
A.~Cheek, L.~Heurtier, Y.~F. Perez-Gonzalez, and J.~Turner, ``{Primordial black
  hole evaporation and dark matter production. II. Interplay with the freeze-in
  or freeze-out mechanism},''
  \href{http://dx.doi.org/10.1103/PhysRevD.105.015023}{{\em Phys. Rev. D}
  {\bfseries 105} no.~1, (2022) 015023},
  \href{http://arxiv.org/abs/2107.00016}{{\ttfamily arXiv:2107.00016
  [hep-ph]}}.

\bibitem{Hooper:2020evu}
D.~Hooper, G.~Krnjaic, J.~March-Russell, S.~D. McDermott, and
  R.~Petrossian-Byrne, ``{Hot Gravitons and Gravitational Waves From Kerr Black
  Holes in the Early Universe},''
  \href{http://arxiv.org/abs/2004.00618}{{\ttfamily arXiv:2004.00618
  [astro-ph.CO]}}.

\bibitem{Arbey:2021ysg}
A.~Arbey, J.~Auffinger, P.~Sandick, B.~Shams Es~Haghi, and K.~Sinha,
  ``{Precision calculation of dark radiation from spinning primordial black
  holes and early matter-dominated eras},''
  \href{http://dx.doi.org/10.1103/PhysRevD.103.123549}{{\em Phys. Rev. D}
  {\bfseries 103} no.~12, (2021) 123549},
  \href{http://arxiv.org/abs/2104.04051}{{\ttfamily arXiv:2104.04051
  [astro-ph.CO]}}.

\bibitem{Sandick:2021gew}
P.~Sandick, B.~S. Es~Haghi, and K.~Sinha, ``{Asymmetric reheating by primordial
  black holes},'' \href{http://dx.doi.org/10.1103/PhysRevD.104.083523}{{\em
  Phys. Rev. D} {\bfseries 104} no.~8, (2021) 083523},
  \href{http://arxiv.org/abs/2108.08329}{{\ttfamily arXiv:2108.08329
  [astro-ph.CO]}}.

\bibitem{Masina:2021zpu}
I.~Masina, ``{Dark Matter and Dark Radiation from Evaporating Kerr Primordial
  Black Holes},'' \href{http://dx.doi.org/10.1134/S0202289321040101}{{\em Grav.
  Cosmol.} {\bfseries 27} no.~4, (2021) 315--330},
  \href{http://arxiv.org/abs/2103.13825}{{\ttfamily arXiv:2103.13825 [gr-qc]}}.

\bibitem{Cheek:2022dbx}
A.~Cheek, L.~Heurtier, Y.~F. Perez-Gonzalez, and J.~Turner, ``{Redshift effects
  in particle production from Kerr primordial black holes},''
  \href{http://dx.doi.org/10.1103/PhysRevD.106.103012}{{\em Phys. Rev. D}
  {\bfseries 106} no.~10, (2022) 103012},
  \href{http://arxiv.org/abs/2207.09462}{{\ttfamily arXiv:2207.09462
  [astro-ph.CO]}}.

\bibitem{Schiavone:2021imu}
F.~Schiavone, D.~Montanino, A.~Mirizzi, and F.~Capozzi, ``{Axion-like particles
  from primordial black holes shining through the Universe},''
  \href{http://dx.doi.org/10.1088/1475-7516/2021/08/063}{{\em JCAP} {\bfseries
  08} (2021) 063}, \href{http://arxiv.org/abs/2107.03420}{{\ttfamily
  arXiv:2107.03420 [hep-ph]}}.

\bibitem{Bernal:2021yyb}
N.~Bernal, F.~Hajkarim, and Y.~Xu, ``{Axion Dark Matter in the Time of
  Primordial Black Holes},''
  \href{http://dx.doi.org/10.1103/PhysRevD.104.075007}{{\em Phys. Rev. D}
  {\bfseries 104} (2021) 075007},
  \href{http://arxiv.org/abs/2107.13575}{{\ttfamily arXiv:2107.13575
  [hep-ph]}}.

\bibitem{Mazde:2022sdx}
K.~Mazde and L.~Visinelli, ``{The Interplay between the Dark Matter Axion and
  Primordial Black Holes},'' \href{http://arxiv.org/abs/2209.14307}{{\ttfamily
  arXiv:2209.14307 [astro-ph.CO]}}.

\bibitem{Li:2022mcf}
H.-J. Li, ``{Primordial black holes induced stochastic axion-photon
  oscillations in primordial magnetic field},''
  \href{http://dx.doi.org/10.1088/1475-7516/2022/11/045}{{\em JCAP} {\bfseries
  11} (2022) 045}, \href{http://arxiv.org/abs/2208.04605}{{\ttfamily
  arXiv:2208.04605 [astro-ph.CO]}}.

\bibitem{Zeldovich:1976vw}
Y.~B. Zeldovich, ``{Charge Asymmetry of the Universe Due to Black Hole
  Evaporation and Weak Interaction Asymmetry},'' {\em Pisma Zh. Eksp. Teor.
  Fiz.} {\bfseries 24} (1976) 29--32.

\bibitem{Carr:1976zz}
B.~J. Carr, ``{Some cosmological consequences of primordial black-hole
  evaporations},'' \href{http://dx.doi.org/10.1086/154351}{{\em Astrophys. J.}
  {\bfseries 206} (1976) 8--25}.

\bibitem{Toussaint:1978br}
D.~Toussaint, S.~B. Treiman, F.~Wilczek, and A.~Zee, ``{Matter - Antimatter
  Accounting, Thermodynamics, and Black Hole Radiation},''
  \href{http://dx.doi.org/10.1103/PhysRevD.19.1036}{{\em Phys. Rev. D}
  {\bfseries 19} (1979) 1036--1045}.

\bibitem{Turner:1979bt}
M.~S. Turner, ``{BARYON PRODUCTION BY PRIMORDIAL BLACK HOLES},''
  \href{http://dx.doi.org/10.1016/0370-2693(79)90095-9}{{\em Phys. Lett. B}
  {\bfseries 89} (1979) 155--159}.

\bibitem{Grillo:1980rt}
A.~F. Grillo, ``{Primordial Black Holes and Baryon Production in Grand Unified
  Theories},'' \href{http://dx.doi.org/10.1016/0370-2693(80)90897-7}{{\em Phys.
  Lett. B} {\bfseries 94} (1980) 364--366}.

\bibitem{Baumann:2007yr}
D.~Baumann, P.~J. Steinhardt, and N.~Turok, ``{Primordial Black Hole
  Baryogenesis},'' \href{http://arxiv.org/abs/hep-th/0703250}{{\ttfamily
  arXiv:hep-th/0703250}}.

\bibitem{Fujita:2014hha}
T.~Fujita, M.~Kawasaki, K.~Harigaya, and R.~Matsuda, ``{Baryon asymmetry, dark
  matter, and density perturbation from primordial black holes},''
  \href{http://dx.doi.org/10.1103/PhysRevD.89.103501}{{\em Phys. Rev. D}
  {\bfseries 89} no.~10, (2014) 103501},
  \href{http://arxiv.org/abs/1401.1909}{{\ttfamily arXiv:1401.1909
  [astro-ph.CO]}}.

\bibitem{Hook:2014mla}
A.~Hook, ``{Baryogenesis from Hawking Radiation},''
  \href{http://dx.doi.org/10.1103/PhysRevD.90.083535}{{\em Phys. Rev. D}
  {\bfseries 90} no.~8, (2014) 083535},
  \href{http://arxiv.org/abs/1404.0113}{{\ttfamily arXiv:1404.0113 [hep-ph]}}.

\bibitem{Hamada:2016jnq}
Y.~Hamada and S.~Iso, ``{Baryon asymmetry from primordial black holes},''
  \href{http://dx.doi.org/10.1093/ptep/ptx011}{{\em PTEP} {\bfseries 2017}
  no.~3, (2017) 033B02}, \href{http://arxiv.org/abs/1610.02586}{{\ttfamily
  arXiv:1610.02586 [hep-ph]}}.

\bibitem{Morrison:2018xla}
L.~Morrison, S.~Profumo, and Y.~Yu, ``{Melanopogenesis: Dark Matter of (almost)
  any Mass and Baryonic Matter from the Evaporation of Primordial Black Holes
  weighing a Ton (or less)},''
  \href{http://dx.doi.org/10.1088/1475-7516/2019/05/005}{{\em JCAP} {\bfseries
  05} (2019) 005}, \href{http://arxiv.org/abs/1812.10606}{{\ttfamily
  arXiv:1812.10606 [astro-ph.CO]}}.

\bibitem{Hooper:2020otu}
D.~Hooper and G.~Krnjaic, ``{GUT Baryogenesis With Primordial Black Holes},''
  \href{http://dx.doi.org/10.1103/PhysRevD.103.043504}{{\em Phys. Rev. D}
  {\bfseries 103} no.~4, (2021) 043504},
  \href{http://arxiv.org/abs/2010.01134}{{\ttfamily arXiv:2010.01134
  [hep-ph]}}.

\bibitem{Bernal:2022pue}
N.~Bernal, C.~S. Fong, Y.~F. Perez-Gonzalez, and J.~Turner, ``{Rescuing
  high-scale leptogenesis using primordial black holes},''
  \href{http://dx.doi.org/10.1103/PhysRevD.106.035019}{{\em Phys. Rev. D}
  {\bfseries 106} no.~3, (2022) 035019},
  \href{http://arxiv.org/abs/2203.08823}{{\ttfamily arXiv:2203.08823
  [hep-ph]}}.

\bibitem{Gehrman:2022imk}
T.~C. Gehrman, B.~Shams Es~Haghi, K.~Sinha, and T.~Xu, ``{Baryogenesis,
  Primordial Black Holes and MHz-GHz Gravitational Waves},''
  \href{http://arxiv.org/abs/2211.08431}{{\ttfamily arXiv:2211.08431
  [hep-ph]}}.

\bibitem{Baker:2021btk}
M.~J. Baker and A.~Thamm, ``{Probing the particle spectrum of nature with
  evaporating black holes},''
  \href{http://dx.doi.org/10.21468/SciPostPhys.12.5.150}{{\em SciPost Phys.}
  {\bfseries 12} no.~5, (2022) 150},
  \href{http://arxiv.org/abs/2105.10506}{{\ttfamily arXiv:2105.10506
  [hep-ph]}}.

\bibitem{Calabrese:2021src}
R.~Calabrese, M.~Chianese, D.~F.~G. Fiorillo, and N.~Saviano, ``{Direct
  detection of light dark matter from evaporating primordial black holes},''
  \href{http://dx.doi.org/10.1103/PhysRevD.105.L021302}{{\em Phys. Rev. D}
  {\bfseries 105} no.~2, (2022) L021302},
  \href{http://arxiv.org/abs/2107.13001}{{\ttfamily arXiv:2107.13001
  [hep-ph]}}.

\bibitem{Calza:2021czr}
M.~Calz\`a, J.~March-Russell, and J.~a.~G. Rosa, ``{Evaporating primordial
  black holes, the string axiverse, and hot dark radiation},''
  \href{http://arxiv.org/abs/2110.13602}{{\ttfamily arXiv:2110.13602
  [astro-ph.CO]}}.

\bibitem{Li:2022jxo}
T.~Li and J.~Liao, ``{Electron-target experiment constraints on light dark
  matter produced in primordial black hole evaporation},''
  \href{http://dx.doi.org/10.1103/PhysRevD.106.055043}{{\em Phys. Rev. D}
  {\bfseries 106} no.~5, (2022) 055043},
  \href{http://arxiv.org/abs/2203.14443}{{\ttfamily arXiv:2203.14443
  [hep-ph]}}.

\bibitem{Calabrese:2022rfa}
R.~Calabrese, M.~Chianese, D.~F.~G. Fiorillo, and N.~Saviano, ``{Electron
  scattering of light new particles from evaporating primordial black holes},''
  \href{http://dx.doi.org/10.1103/PhysRevD.105.103024}{{\em Phys. Rev. D}
  {\bfseries 105} no.~10, (2022) 103024},
  \href{http://arxiv.org/abs/2203.17093}{{\ttfamily arXiv:2203.17093
  [hep-ph]}}.

\bibitem{Li:2022xqh}
T.~Li and R.-J. Zhang, ``{Axionlike particles from primordial black hole
  evaporation and their detection in neutrino experiments},''
  \href{http://dx.doi.org/10.1103/PhysRevD.106.095034}{{\em Phys. Rev. D}
  {\bfseries 106} no.~9, (2022) 095034},
  \href{http://arxiv.org/abs/2208.02696}{{\ttfamily arXiv:2208.02696
  [hep-ph]}}.

\bibitem{Baker:2022rkn}
M.~J. Baker and A.~Thamm, ``{Black Hole Evaporation Beyond the Standard Model
  of Particle Physics},'' \href{http://arxiv.org/abs/2210.02805}{{\ttfamily
  arXiv:2210.02805 [hep-ph]}}.

\bibitem{Kappadath:1998PhDT}
S.~C. {Kappadath}, {\em {Measurement of the Cosmic Diffuse Gamma-Ray Spectrum
  from 800 KEV to 30 Mev}}.
\newblock PhD thesis, University of New Hampshire, United States, Jan., 1998.

\bibitem{Fermi-LAT:2009ihh}
{\bfseries Fermi-LAT} Collaboration, W.~B. Atwood {\em et~al.}, ``{The Large
  Area Telescope on the Fermi Gamma-ray Space Telescope Mission},''
  \href{http://dx.doi.org/10.1088/0004-637X/697/2/1071}{{\em Astrophys. J.}
  {\bfseries 697} (2009) 1071--1102},
  \href{http://arxiv.org/abs/0902.1089}{{\ttfamily arXiv:0902.1089
  [astro-ph.IM]}}.

\bibitem{PhysRevLett.38.1440}
R.~D. Peccei and H.~R. Quinn, ``$\mathrm{CP}$ conservation in the presence of
  pseudoparticles,'' \href{http://dx.doi.org/10.1103/PhysRevLett.38.1440}{{\em
  Phys. Rev. Lett.} {\bfseries 38} (Jun, 1977) 1440--1443}.
  \url{https://link.aps.org/doi/10.1103/PhysRevLett.38.1440}.

\bibitem{PhysRevD.16.1791}
R.~D. Peccei and H.~R. Quinn, ``Constraints imposed by $\mathrm{CP}$
  conservation in the presence of pseudoparticles,''
  \href{http://dx.doi.org/10.1103/PhysRevD.16.1791}{{\em Phys. Rev. D}
  {\bfseries 16} (Sep, 1977) 1791--1797}.
  \url{https://link.aps.org/doi/10.1103/PhysRevD.16.1791}.

\bibitem{ABBOTT1983133}
L.~Abbott and P.~Sikivie, ``A cosmological bound on the invisible axion,''
  \href{http://dx.doi.org/https://doi.org/10.1016/0370-2693(83)90638-X}{{\em
  Physics Letters B} {\bfseries 120} no.~1, (1983) 133--136}.
  \url{https://www.sciencedirect.com/science/article/pii/037026938390638X}.

\bibitem{PRESKILL1983127}
J.~Preskill, M.~B. Wise, and F.~Wilczek, ``Cosmology of the invisible axion,''
  \href{http://dx.doi.org/https://doi.org/10.1016/0370-2693(83)90637-8}{{\em
  Physics Letters B} {\bfseries 120} no.~1, (1983) 127--132}.
  \url{https://www.sciencedirect.com/science/article/pii/0370269383906378}.

\bibitem{DINE1983137}
M.~Dine and W.~Fischler, ``The not-so-harmless axion,''
  \href{http://dx.doi.org/https://doi.org/10.1016/0370-2693(83)90639-1}{{\em
  Physics Letters B} {\bfseries 120} no.~1, (1983) 137--141}.
  \url{https://www.sciencedirect.com/science/article/pii/0370269383906391}.

\bibitem{Co:2017mop}
R.~T. Co, L.~J. Hall, and K.~Harigaya, ``{QCD Axion Dark Matter with a Small
  Decay Constant},''
  \href{http://dx.doi.org/10.1103/PhysRevLett.120.211602}{{\em Phys. Rev.
  Lett.} {\bfseries 120} no.~21, (2018) 211602},
  \href{http://arxiv.org/abs/1711.10486}{{\ttfamily arXiv:1711.10486
  [hep-ph]}}.

\bibitem{Freese:1990rb}
K.~Freese, J.~A. Frieman, and A.~V. Olinto, ``{Natural inflation with pseudo -
  Nambu-Goldstone bosons},''
  \href{http://dx.doi.org/10.1103/PhysRevLett.65.3233}{{\em Phys. Rev. Lett.}
  {\bfseries 65} (1990) 3233--3236}.

\bibitem{Kim:2004rp}
J.~E. Kim, H.~P. Nilles, and M.~Peloso, ``{Completing natural inflation},''
  \href{http://dx.doi.org/10.1088/1475-7516/2005/01/005}{{\em JCAP} {\bfseries
  01} (2005) 005}, \href{http://arxiv.org/abs/hep-ph/0409138}{{\ttfamily
  arXiv:hep-ph/0409138}}.

\bibitem{Pajer:2013fsa}
E.~Pajer and M.~Peloso, ``{A review of Axion Inflation in the era of Planck},''
  \href{http://dx.doi.org/10.1088/0264-9381/30/21/214002}{{\em Class. Quant.
  Grav.} {\bfseries 30} (2013) 214002},
  \href{http://arxiv.org/abs/1305.3557}{{\ttfamily arXiv:1305.3557 [hep-th]}}.

\bibitem{Graham:2015cka}
P.~W. Graham, D.~E. Kaplan, and S.~Rajendran, ``{Cosmological Relaxation of the
  Electroweak Scale},''
  \href{http://dx.doi.org/10.1103/PhysRevLett.115.221801}{{\em Phys. Rev.
  Lett.} {\bfseries 115} no.~22, (2015) 221801},
  \href{http://arxiv.org/abs/1504.07551}{{\ttfamily arXiv:1504.07551
  [hep-ph]}}.

\bibitem{Gupta:2015uea}
R.~S. Gupta, Z.~Komargodski, G.~Perez, and L.~Ubaldi, ``{Is the Relaxion an
  Axion?},'' \href{http://dx.doi.org/10.1007/JHEP02(2016)166}{{\em JHEP}
  {\bfseries 02} (2016) 166}, \href{http://arxiv.org/abs/1509.00047}{{\ttfamily
  arXiv:1509.00047 [hep-ph]}}.

\bibitem{Choi:2015fiu}
K.~Choi and S.~H. Im, ``{Realizing the relaxion from multiple axions and its UV
  completion with high scale supersymmetry},''
  \href{http://dx.doi.org/10.1007/JHEP01(2016)149}{{\em JHEP} {\bfseries 01}
  (2016) 149}, \href{http://arxiv.org/abs/1511.00132}{{\ttfamily
  arXiv:1511.00132 [hep-ph]}}.

\bibitem{Sikivie:2020zpn}
P.~Sikivie, ``{Invisible Axion Search Methods},''
  \href{http://dx.doi.org/10.1103/RevModPhys.93.015004}{{\em Rev. Mod. Phys.}
  {\bfseries 93} no.~1, (2021) 015004},
  \href{http://arxiv.org/abs/2003.02206}{{\ttfamily arXiv:2003.02206
  [hep-ph]}}.

\bibitem{Bauer:2017ris}
M.~Bauer, M.~Neubert, and A.~Thamm, ``{Collider Probes of Axion-Like
  Particles},'' \href{http://dx.doi.org/10.1007/JHEP12(2017)044}{{\em JHEP}
  {\bfseries 12} (2017) 044}, \href{http://arxiv.org/abs/1708.00443}{{\ttfamily
  arXiv:1708.00443 [hep-ph]}}.

\bibitem{Bauer:2018uxu}
M.~Bauer, M.~Heiles, M.~Neubert, and A.~Thamm, ``{Axion-Like Particles at
  Future Colliders},''
  \href{http://dx.doi.org/10.1140/epjc/s10052-019-6587-9}{{\em Eur. Phys. J. C}
  {\bfseries 79} no.~1, (2019) 74},
  \href{http://arxiv.org/abs/1808.10323}{{\ttfamily arXiv:1808.10323
  [hep-ph]}}.

\bibitem{Adams:2022pbo}
C.~B. Adams {\em et~al.}, ``{Axion Dark Matter},'' in {\em {2022 Snowmass
  Summer Study}}.
\newblock 3, 2022.
\newblock \href{http://arxiv.org/abs/2203.14923}{{\ttfamily arXiv:2203.14923
  [hep-ex]}}.

\bibitem{Irastorza:2018dyq}
I.~G. Irastorza and J.~Redondo, ``{New experimental approaches in the search
  for axion-like particles},''
  \href{http://dx.doi.org/10.1016/j.ppnp.2018.05.003}{{\em Prog. Part. Nucl.
  Phys.} {\bfseries 102} (2018) 89--159},
  \href{http://arxiv.org/abs/1801.08127}{{\ttfamily arXiv:1801.08127
  [hep-ph]}}.

\bibitem{DiLuzio:2020wdo}
L.~Di~Luzio, M.~Giannotti, E.~Nardi, and L.~Visinelli, ``{The landscape of QCD
  axion models},'' \href{http://dx.doi.org/10.1016/j.physrep.2020.06.002}{{\em
  Phys. Rept.} {\bfseries 870} (2020) 1--117},
  \href{http://arxiv.org/abs/2003.01100}{{\ttfamily arXiv:2003.01100
  [hep-ph]}}.

\bibitem{Choi:2020rgn}
K.~Choi, S.~H. Im, and C.~Sub~Shin, ``{Recent Progress in the Physics of Axions
  and Axion-Like Particles},''
  \href{http://dx.doi.org/10.1146/annurev-nucl-120720-031147}{{\em Ann. Rev.
  Nucl. Part. Sci.} {\bfseries 71} (2021) 225--252},
  \href{http://arxiv.org/abs/2012.05029}{{\ttfamily arXiv:2012.05029
  [hep-ph]}}.

\bibitem{AxionLimits}
C.~O'Hare, ``cajohare/axionlimits: Axionlimits.''
  \url{https://cajohare.github.io/AxionLimits/}, July, 2020.

\bibitem{Ayala:2014pea}
A.~Ayala, I.~Dom\'\i{}nguez, M.~Giannotti, A.~Mirizzi, and O.~Straniero,
  ``{Revisiting the bound on axion-photon coupling from Globular Clusters},''
  \href{http://dx.doi.org/10.1103/PhysRevLett.113.191302}{{\em Phys. Rev.
  Lett.} {\bfseries 113} no.~19, (2014) 191302},
  \href{http://arxiv.org/abs/1406.6053}{{\ttfamily arXiv:1406.6053
  [astro-ph.SR]}}.

\bibitem{2015JCAP...10..015V}
N.~{Vinyoles}, A.~{Serenelli}, F.~L. {Villante}, S.~{Basu}, J.~{Redondo}, and
  J.~{Isern}, ``{New axion and hidden photon constraints from a solar data
  global fit},'' \href{http://dx.doi.org/10.1088/1475-7516/2015/10/015}{{\em
  \jcap} {\bfseries 2015} no.~10, (Oct., 2015) 015--015},
  \href{http://arxiv.org/abs/1501.01639}{{\ttfamily arXiv:1501.01639
  [astro-ph.SR]}}.

\bibitem{Knapen:2016moh}
S.~Knapen, T.~Lin, H.~K. Lou, and T.~Melia, ``{Searching for Axionlike
  Particles with Ultraperipheral Heavy-Ion Collisions},''
  \href{http://dx.doi.org/10.1103/PhysRevLett.118.171801}{{\em Phys. Rev.
  Lett.} {\bfseries 118} no.~17, (2017) 171801},
  \href{http://arxiv.org/abs/1607.06083}{{\ttfamily arXiv:1607.06083
  [hep-ph]}}.

\bibitem{Belle-II:2020jti}
{\bfseries Belle-II} Collaboration, F.~Abudin\'en {\em et~al.}, ``{Search for
  Axion-Like Particles produced in $e^+e^-$ collisions at Belle II},''
  \href{http://dx.doi.org/10.1103/PhysRevLett.125.161806}{{\em Phys. Rev.
  Lett.} {\bfseries 125} no.~16, (2020) 161806},
  \href{http://arxiv.org/abs/2007.13071}{{\ttfamily arXiv:2007.13071
  [hep-ex]}}.

\bibitem{Lucente:2020whw}
G.~Lucente, P.~Carenza, T.~Fischer, M.~Giannotti, and A.~Mirizzi, ``{Heavy
  axion-like particles and core-collapse supernovae: constraints and impact on
  the explosion mechanism},''
  \href{http://dx.doi.org/10.1088/1475-7516/2020/12/008}{{\em JCAP} {\bfseries
  12} (2020) 008}, \href{http://arxiv.org/abs/2008.04918}{{\ttfamily
  arXiv:2008.04918 [hep-ph]}}.

\bibitem{Dolan:2021rya}
M.~J. Dolan, F.~J. Hiskens, and R.~R. Volkas, ``{Constraining axion-like
  particles using the white dwarf initial-final mass relation},''
  \href{http://dx.doi.org/10.1088/1475-7516/2021/09/010}{{\em JCAP} {\bfseries
  09} (2021) 010}, \href{http://arxiv.org/abs/2102.00379}{{\ttfamily
  arXiv:2102.00379 [hep-ph]}}.

\bibitem{Dessert:2021bkv}
C.~Dessert, A.~J. Long, and B.~R. Safdi, ``{No Evidence for Axions from Chandra
  Observation of the Magnetic White Dwarf RE J0317-853},''
  \href{http://dx.doi.org/10.1103/PhysRevLett.128.071102}{{\em Phys. Rev.
  Lett.} {\bfseries 128} no.~7, (2022) 071102},
  \href{http://arxiv.org/abs/2104.12772}{{\ttfamily arXiv:2104.12772
  [hep-ph]}}.

\bibitem{Caputo:2022mah}
A.~Caputo, H.-T. Janka, G.~Raffelt, and E.~Vitagliano, ``{Low-Energy Supernovae
  Severely Constrain Radiative Particle Decays},''
  \href{http://dx.doi.org/10.1103/PhysRevLett.128.221103}{{\em Phys. Rev.
  Lett.} {\bfseries 128} no.~22, (2022) 221103},
  \href{http://arxiv.org/abs/2201.09890}{{\ttfamily arXiv:2201.09890
  [astro-ph.HE]}}.

\bibitem{Dessert:2022yqq}
C.~Dessert, D.~Dunsky, and B.~R. Safdi, ``{Upper limit on the axion-photon
  coupling from magnetic white dwarf polarization},''
  \href{http://dx.doi.org/10.1103/PhysRevD.105.103034}{{\em Phys. Rev. D}
  {\bfseries 105} no.~10, (2022) 103034},
  \href{http://arxiv.org/abs/2203.04319}{{\ttfamily arXiv:2203.04319
  [hep-ph]}}.

\bibitem{Kling:2022ehv}
F.~Kling and P.~Qu\'\i{}lez, ``{ALP searches at the LHC: FASER as a
  light-shining-through-walls experiment},''
  \href{http://dx.doi.org/10.1103/PhysRevD.106.055036}{{\em Phys. Rev. D}
  {\bfseries 106} no.~5, (2022) 055036},
  \href{http://arxiv.org/abs/2204.03599}{{\ttfamily arXiv:2204.03599
  [hep-ph]}}.

\bibitem{DeRocco:2022jyq}
W.~DeRocco, S.~Wegsman, B.~Grefenstette, J.~Huang, and K.~Van~Tilburg, ``{First
  Indirect Detection Constraints on Axions in the Solar Basin},''
  \href{http://dx.doi.org/10.1103/PhysRevLett.129.101101}{{\em Phys. Rev.
  Lett.} {\bfseries 129} no.~10, (2022) 101101},
  \href{http://arxiv.org/abs/2205.05700}{{\ttfamily arXiv:2205.05700
  [hep-ph]}}.

\bibitem{Balazs:2022tjl}
C.~Bal\'azs {\em et~al.}, ``{Cosmological constraints on decaying axion-like
  particles: a global analysis},''
  \href{http://arxiv.org/abs/2205.13549}{{\ttfamily arXiv:2205.13549
  [astro-ph.CO]}}.

\bibitem{Dolan:2022kul}
M.~J. Dolan, F.~J. Hiskens, and R.~R. Volkas, ``{Advancing globular cluster
  constraints on the axion-photon coupling},''
  \href{http://dx.doi.org/10.1088/1475-7516/2022/10/096}{{\em JCAP} {\bfseries
  10} (2022) 096}, \href{http://arxiv.org/abs/2207.03102}{{\ttfamily
  arXiv:2207.03102 [hep-ph]}}.

\bibitem{Langhoff:2022bij}
K.~Langhoff, N.~J. Outmezguine, and N.~L. Rodd, ``{Irreducible Axion
  Background},'' \href{http://dx.doi.org/10.1103/PhysRevLett.129.241101}{{\em
  Phys. Rev. Lett.} {\bfseries 129} no.~24, (2022) 241101},
  \href{http://arxiv.org/abs/2209.06216}{{\ttfamily arXiv:2209.06216
  [hep-ph]}}.

\bibitem{BESIII:2022rzz}
{\bfseries BESIII} Collaboration, M.~Ablikim {\em et~al.}, ``{Search for an
  axion-like particle in radiative J/\ensuremath{\psi} decays},''
  \href{http://dx.doi.org/10.1016/j.physletb.2023.137698}{{\em Phys. Lett. B}
  {\bfseries 838} (2023) 137698},
  \href{http://arxiv.org/abs/2211.12699}{{\ttfamily arXiv:2211.12699
  [hep-ex]}}.

\bibitem{SREDNICKI1985689}
M.~Srednicki, ``Axion couplings to matter: (i). cp-conserving parts,''
  \href{http://dx.doi.org/https://doi.org/10.1016/0550-3213(85)90054-9}{{\em
  Nuclear Physics B} {\bfseries 260} no.~3, (1985) 689--700}.
  \url{https://www.sciencedirect.com/science/article/pii/0550321385900549}.

\bibitem{Kim:2008hd}
J.~E. Kim and G.~Carosi, ``{Axions and the Strong CP Problem},''
  \href{http://dx.doi.org/10.1103/RevModPhys.82.557}{{\em Rev. Mod. Phys.}
  {\bfseries 82} (2010) 557--602},
  \href{http://arxiv.org/abs/0807.3125}{{\ttfamily arXiv:0807.3125 [hep-ph]}}.
  [Erratum: Rev.Mod.Phys. 91, 049902 (2019)].

\bibitem{Agashe:2022jgk}
K.~Agashe, J.~H. Chang, S.~J. Clark, B.~Dutta, Y.~Tsai, and T.~Xu,
  ``{Correlating gravitational wave and gamma-ray signals from primordial black
  holes},'' \href{http://dx.doi.org/10.1103/PhysRevD.105.123009}{{\em Phys.
  Rev. D} {\bfseries 105} no.~12, (2022) 123009},
  \href{http://arxiv.org/abs/2202.04653}{{\ttfamily arXiv:2202.04653
  [astro-ph.CO]}}.

\bibitem{Depta:2020wmr}
P.~F. Depta, M.~Hufnagel, and K.~Schmidt-Hoberg, ``{Robust cosmological
  constraints on axion-like particles},''
  \href{http://dx.doi.org/10.1088/1475-7516/2020/05/009}{{\em JCAP} {\bfseries
  05} (2020) 009}, \href{http://arxiv.org/abs/2002.08370}{{\ttfamily
  arXiv:2002.08370 [hep-ph]}}.

\bibitem{Depta:2020zbh}
P.~F. Depta, M.~Hufnagel, and K.~Schmidt-Hoberg, ``{Updated BBN constraints on
  electromagnetic decays of MeV-scale particles},''
  \href{http://dx.doi.org/10.1088/1475-7516/2021/04/011}{{\em JCAP} {\bfseries
  04} (2021) 011}, \href{http://arxiv.org/abs/2011.06519}{{\ttfamily
  arXiv:2011.06519 [hep-ph]}}.

\bibitem{Caputo:2021rux}
A.~Caputo, G.~Raffelt, and E.~Vitagliano, ``{Muonic boson limits: Supernova
  redux},'' \href{http://dx.doi.org/10.1103/PhysRevD.105.035022}{{\em Phys.
  Rev. D} {\bfseries 105} no.~3, (2022) 035022},
  \href{http://arxiv.org/abs/2109.03244}{{\ttfamily arXiv:2109.03244
  [hep-ph]}}.

\bibitem{Hawking:1974rv}
S.~W. Hawking, ``{Black hole explosions},''
  \href{http://dx.doi.org/10.1038/248030a0}{{\em Nature} {\bfseries 248} (1974)
  30--31}.

\bibitem{Page:1976df}
D.~N. Page, ``{Particle Emission Rates from a Black Hole: Massless Particles
  from an Uncharged, Nonrotating Hole},''
  \href{http://dx.doi.org/10.1103/PhysRevD.13.198}{{\em Phys. Rev. D}
  {\bfseries 13} (1976) 198--206}.

\bibitem{MacGibbon:1990zk}
J.~H. MacGibbon and B.~R. Webber, ``{Quark and gluon jet emission from
  primordial black holes: The instantaneous spectra},''
  \href{http://dx.doi.org/10.1103/PhysRevD.41.3052}{{\em Phys. Rev. D}
  {\bfseries 41} (1990) 3052--3079}.

\bibitem{Arbey:2019mbc}
A.~Arbey and J.~Auffinger, ``{BlackHawk: A public code for calculating the
  Hawking evaporation spectra of any black hole distribution},''
  \href{http://dx.doi.org/10.1140/epjc/s10052-019-7161-1}{{\em Eur. Phys. J. C}
  {\bfseries 79} no.~8, (2019) 693},
  \href{http://arxiv.org/abs/1905.04268}{{\ttfamily arXiv:1905.04268 [gr-qc]}}.

\bibitem{Arbey:2021mbl}
A.~Arbey and J.~Auffinger, ``{Physics Beyond the Standard Model with BlackHawk
  v2.0},'' \href{http://dx.doi.org/10.1140/epjc/s10052-021-09702-8}{{\em Eur.
  Phys. J. C} {\bfseries 81} (2021) 910},
  \href{http://arxiv.org/abs/2108.02737}{{\ttfamily arXiv:2108.02737 [gr-qc]}}.

\bibitem{Navarro:1996gj}
J.~F. Navarro, C.~S. Frenk, and S.~D.~M. White, ``{A Universal density profile
  from hierarchical clustering},'' \href{http://dx.doi.org/10.1086/304888}{{\em
  Astrophys. J.} {\bfseries 490} (1997) 493--508},
  \href{http://arxiv.org/abs/astro-ph/9611107}{{\ttfamily
  arXiv:astro-ph/9611107}}.

\bibitem{2019JCAP...10..037D}
P.~F. {de Salas}, K.~{Malhan}, K.~{Freese}, K.~{Hattori}, and M.~{Valluri},
  ``{On the estimation of the local dark matter density using the rotation
  curve of the Milky Way},''
  \href{http://dx.doi.org/10.1088/1475-7516/2019/10/037}{{\em \jcap} {\bfseries
  2019} no.~10, (Oct., 2019) 037},
  \href{http://arxiv.org/abs/1906.06133}{{\ttfamily arXiv:1906.06133
  [astro-ph.GA]}}.

\bibitem{Flores:2021jas}
M.~M. Flores and A.~Kusenko, ``{Primordial black holes as a dark matter
  candidate in theories with supersymmetry and inflation},''
  \href{http://arxiv.org/abs/2108.08416}{{\ttfamily arXiv:2108.08416
  [hep-ph]}}.

\bibitem{Jung:2021mku}
T.~H. Jung and T.~Okui, ``{Primordial black holes from bubble collisions during
  a first-order phase transition},''
  \href{http://arxiv.org/abs/2110.04271}{{\ttfamily arXiv:2110.04271
  [hep-ph]}}.

\bibitem{Cowan:2010js}
G.~Cowan, K.~Cranmer, E.~Gross, and O.~Vitells, ``{Asymptotic formulae for
  likelihood-based tests of new physics},''
  \href{http://dx.doi.org/10.1140/epjc/s10052-011-1554-0}{{\em Eur. Phys. J. C}
  {\bfseries 71} (2011) 1554}, \href{http://arxiv.org/abs/1007.1727}{{\ttfamily
  arXiv:1007.1727 [physics.data-an]}}. [Erratum: Eur.Phys.J.C 73, 2501 (2013)].

\bibitem{Rolke:2004mj}
W.~A. Rolke, A.~M. Lopez, and J.~Conrad, ``{Limits and confidence intervals in
  the presence of nuisance parameters},''
  \href{http://dx.doi.org/10.1016/j.nima.2005.05.068}{{\em Nucl. Instrum. Meth.
  A} {\bfseries 551} (2005) 493--503},
  \href{http://arxiv.org/abs/physics/0403059}{{\ttfamily
  arXiv:physics/0403059}}.

\bibitem{Bringmann:2012vr}
T.~Bringmann, X.~Huang, A.~Ibarra, S.~Vogl, and C.~Weniger, ``{Fermi LAT Search
  for Internal Bremsstrahlung Signatures from Dark Matter Annihilation},''
  \href{http://dx.doi.org/10.1088/1475-7516/2012/07/054}{{\em JCAP} {\bfseries
  07} (2012) 054}, \href{http://arxiv.org/abs/1203.1312}{{\ttfamily
  arXiv:1203.1312 [hep-ph]}}.

\bibitem{Fermi-LAT:2015kyq}
{\bfseries Fermi-LAT} Collaboration, M.~Ackermann {\em et~al.}, ``{Updated
  search for spectral lines from Galactic dark matter interactions with pass 8
  data from the Fermi Large Area Telescope},''
  \href{http://dx.doi.org/10.1103/PhysRevD.91.122002}{{\em Phys. Rev. D}
  {\bfseries 91} no.~12, (2015) 122002},
  \href{http://arxiv.org/abs/1506.00013}{{\ttfamily arXiv:1506.00013
  [astro-ph.HE]}}.

\bibitem{Clark:2018ghm}
S.~Clark, B.~Dutta, Y.~Gao, Y.-Z. Ma, and L.~E. Strigari, ``{21 cm limits on
  decaying dark matter and primordial black holes},''
  \href{http://dx.doi.org/10.1103/PhysRevD.98.043006}{{\em Phys. Rev. D}
  {\bfseries 98} no.~4, (2018) 043006},
  \href{http://arxiv.org/abs/1803.09390}{{\ttfamily arXiv:1803.09390
  [astro-ph.HE]}}.

\bibitem{Boudaud:2018hqb}
M.~Boudaud and M.~Cirelli, ``{Voyager 1 $e^\pm$ Further Constrain Primordial
  Black Holes as Dark Matter},''
  \href{http://dx.doi.org/10.1103/PhysRevLett.122.041104}{{\em Phys. Rev.
  Lett.} {\bfseries 122} no.~4, (2019) 041104},
  \href{http://arxiv.org/abs/1807.03075}{{\ttfamily arXiv:1807.03075
  [astro-ph.HE]}}.

\bibitem{Poulin:2016anj}
V.~Poulin, J.~Lesgourgues, and P.~D. Serpico, ``{Cosmological constraints on
  exotic injection of electromagnetic energy},''
  \href{http://dx.doi.org/10.1088/1475-7516/2017/03/043}{{\em JCAP} {\bfseries
  03} (2017) 043}, \href{http://arxiv.org/abs/1610.10051}{{\ttfamily
  arXiv:1610.10051 [astro-ph.CO]}}.

\bibitem{Clark:2016nst}
S.~Clark, B.~Dutta, Y.~Gao, L.~E. Strigari, and S.~Watson, ``{Planck Constraint
  on Relic Primordial Black Holes},''
  \href{http://dx.doi.org/10.1103/PhysRevD.95.083006}{{\em Phys. Rev. D}
  {\bfseries 95} no.~8, (2017) 083006},
  \href{http://arxiv.org/abs/1612.07738}{{\ttfamily arXiv:1612.07738
  [astro-ph.CO]}}.

\bibitem{DeRocco:2019fjq}
W.~DeRocco and P.~W. Graham, ``{Constraining Primordial Black Hole Abundance
  with the Galactic 511 keV Line},''
  \href{http://dx.doi.org/10.1103/PhysRevLett.123.251102}{{\em Phys. Rev.
  Lett.} {\bfseries 123} no.~25, (2019) 251102},
  \href{http://arxiv.org/abs/1906.07740}{{\ttfamily arXiv:1906.07740
  [astro-ph.CO]}}.

\bibitem{Kim:2020ngi}
H.~Kim, ``{A constraint on light primordial black holes from the interstellar
  medium temperature},'' \href{http://dx.doi.org/10.1093/mnras/stab1222}{{\em
  Mon. Not. Roy. Astron. Soc.} {\bfseries 504} no.~4, (2021) 5475--5484},
  \href{http://arxiv.org/abs/2007.07739}{{\ttfamily arXiv:2007.07739
  [hep-ph]}}.

\bibitem{Saha:2021pqf}
A.~K. Saha and R.~Laha, ``{Sensitivities on nonspinning and spinning primordial
  black hole dark matter with global 21-cm troughs},''
  \href{http://dx.doi.org/10.1103/PhysRevD.105.103026}{{\em Phys. Rev. D}
  {\bfseries 105} no.~10, (2022) 103026},
  \href{http://arxiv.org/abs/2112.10794}{{\ttfamily arXiv:2112.10794
  [astro-ph.CO]}}.

\bibitem{Green:2020jor}
A.~M. Green and B.~J. Kavanagh, ``{Primordial Black Holes as a dark matter
  candidate},'' \href{http://dx.doi.org/10.1088/1361-6471/abc534}{{\em J. Phys.
  G} {\bfseries 48} no.~4, (2021) 043001},
  \href{http://arxiv.org/abs/2007.10722}{{\ttfamily arXiv:2007.10722
  [astro-ph.CO]}}.

\bibitem{Jho:2022wxd}
Y.~Jho, T.-G. Kim, J.-C. Park, S.~C. Park, and Y.~Park, ``{Axions from
  Primordial Black Holes},'' \href{http://arxiv.org/abs/2212.11977}{{\ttfamily
  arXiv:2212.11977 [hep-ph]}}.

\end{thebibliography}\endgroup

\end{document}